%% file: gpce22.tex
\documentclass[sigplan,screen]{acmart}
\usepackage[normalem]{ulem}
\usepackage{proof}
\usepackage{tikz-cd}
\usepackage{balance}

\AtBeginDocument{%
  \providecommand\BibTeX{{%
    \normalfont B\kern-0.5em{\scshape i\kern-0.25em b}\kern-0.8em\TeX}}}

\setcopyright{acmcopyright}
\acmPrice{15.00}
\acmDOI{10.1145/3564719.3568691}
\acmYear{2022}
\copyrightyear{2022}
\acmSubmissionID{splashws22gpcemain-p12-p}
\acmISBN{978-1-4503-9920-3/22/12}
\acmConference[GPCE '22]{Proceedings of the 21st ACM SIGPLAN International Conference on Generative Programming: Concepts and Experiences}{December 06--07, 2022}{Auckland, New Zealand}
\acmBooktitle{Proceedings of the 21st ACM SIGPLAN International Conference on Generative Programming: Concepts and Experiences (GPCE '22), December 06--07, 2022, Auckland, New Zealand}
\received{2022-08-12}
\received[accepted]{2022-10-10}

\input{command.tex}

\input{math.tex}

\begin{document}

\title{Type System for Four Delimited Control Operators}

\author{Chiaki Ishio}
\email{ishio.chiaki@is.ocha.ac.jp}
\affiliation{%
  \institution{Ochanomizu University}
  \state{Tokyo}
  \country{Japan}
  \postcode{1128610}
}

\author{Kenichi Asai}
\email{asai@is.ocha.ac.jp}
\affiliation{%
  \institution{Ochanomizu University}
  \state{Tokyo}
  \country{Japan}
  \postcode{1128610}
}


\input{abstract}



\maketitle

\input{intro}
\input{delimited-control}
\input{ds-cps}
\input{type-system}
\input{other-systems}
\input{other-SRZ}
\input{related}

\input{conclusion}

\begin{acks}
We thank Youyou Cong and the anonymous reviewers for their valuable comments and feedback.
This work was partly supported by JSPS KAKENHI under Grant No.~JP22H03563.
\end{acks}

\appendix
\input{appendix}

\balance
\bibliographystyle{ACM-Reference-Format}
\bibliography{bibmaster.bib}

\end{document}

%% file: command.tex
\newcommand{\Rmath}[1]{\text{\color{red}{$#1$}}}
\newcommand{\Bmath}[1]{\text{\color{blue}{$#1$}}}
\newcommand{\Wmath}[1]{\text{\color{white}{$#1$}}}

\newcommand{\Rtext}[1]{\textcolor{red}{#1}}

\newcommand\todo[1]{\textbf{To Do:~{#1}}}

\newcommand\Fig[1]{Figure~{#1}}

\newcommand\Sec[1]{Section~{#1}}

\newcommand\Thm[1]{Theorem~{#1}}

\newcommand\ttReset{\texttt{reset}}
\newcommand\ttS{\texttt{shift}}
\newcommand\ttR{\texttt{reset}}
\newcommand\ttSR{\texttt{shift/reset}}
\newcommand\ttSZ{\texttt{shift0}}
\newcommand\ttRZ{\texttt{reset0}}
\newcommand\ttSRZ{\texttt{shift0/reset0}}

\newcommand\ttC{\texttt{control}}
\newcommand\ttP{\texttt{prompt}}
\newcommand\ttCP{\texttt{control/prompt}}
\newcommand\ttCZ{\texttt{control0}}
\newcommand\ttPZ{\texttt{prompt0}}
\newcommand\ttCPZ{\texttt{control0/prompt0}}

\newcommand\sffD{\textsf{4D}}
\newcommand\sffDfun{\textsf{4D}_{\textsf{fun}}}
\newcommand\sfMB{\textsf{MB}}
\newcommand\sfDF{\textsf{DF}}
\newcommand\sfDFtwo{\textsf{DF}_{\textsf{2}}}

\newcommand\sfCP{\textsf{CP}}

\newcommand\ld{$\lambda_{D}$}
\newcommand\lc{$\lambda_{C}$}
\newcommand\lsz{$\lambda_{S_0}$}
\newcommand\Fpp[2]{^{#1}\mathcal{F}^{#2}} 

\def\maru#1{{\ooalign{\hfil
  \ifnum#1>999 \resizebox{.25\width}{\height}{#1}\else%
  \ifnum#1>99 \resizebox{.33\width}{\height}{#1}\else%
  \ifnum#1>9 \resizebox{.5\width}{\height}{#1}\else #1%
  \fi\fi\fi%
\/\hfil\crcr%
\raise.167ex\hbox{\mathhexbox20D}}}}

\newcommand\CaseOf[5]{\texttt{case}\,{#1}\,\texttt{of}\,{#2}\Rightarrow {#3}\,|\,{#4}\Rightarrow {#5}}
\newcommand\TypTrailMc[7]{{#1}\,\langle {#2},\,{#3} \rangle \,{#4}\,\langle {#5},\,{#6}\rangle\,{#7}}
\newcommand\Arrow[4]{{#1}\rightarrow \langle {#2},\,{#3} \rangle\,{#4}}
\newcommand\ArrowMB[3]{{#1}\xrightarrow{#2}{#3}}
\newcommand\AnnMB[4]{\,[{#1}\,{#2}]\,{#3}\,{#4}}

\newcommand\toMBAnn{\leftharpoonup^{\sigma}_{\textsf{D}'}}
\newcommand\toMBT[1]{{#1}^{\leftharpoonup^{\tau}_{\textsf{D}'}}}
\newcommand\toMBA[2]{({#1},\,{#2})^{\leftharpoonup^{{\sigma}}_{\textsf{D}'}}}
\newcommand\toMBG[1]{{#1}^{\leftharpoonup^{{\Gamma}}_{\textsf{D}'}}}

\newcommand\toFDAnn{\rightharpoonup^{\sigma}_{\textsf{D}'}}
\newcommand\toFDT[1]{{#1}^{\rightharpoonup^{\tau}_{\textsf{D}'}}}
\newcommand\toFDG[1]{{#1}^{\rightharpoonup^{\Gamma}_{\textsf{D}'}}}
\newcommand\toFDA[2]{({#1},\,{#2})^{\rightharpoonup^{{\sigma}}_{\textsf{D}'}}}
\newcommand\fDdashfDsuper[1]{\rightharpoonup^{#1}_{\textsf{D}}}
\newcommand\fDdashfD[1]{{#1}^{\fDdashfDsuper{\tau}}}
\newcommand\fDdashfDmeta[1]{{#1}^{\fDdashfDsuper{\sigma}}}
\newcommand\fDdashfDgamma[1]{{#1}^{\fDdashfDsuper{\Gamma}}}
\newcommand\fDfDdashsuper[1]{\leftharpoonup^{#1}_{\textsf{D}}}
\newcommand\fDfDdash[1]{{#1}^{\fDfDdashsuper{\tau}}}
\newcommand\fDfDdashmeta[1]{{#1}^{\fDfDdashsuper{\sigma}}}
\newcommand\fDfDdashgamma[1]{{#1}^{\fDfDdashsuper{\Gamma}}}
\newcommand\DFonetwo[2]{({#1},\,{#2})^{\rightharpoonup^\tau_\textsf{A}}}
\newcommand\DFonetwogamma[1]{{#1}^{\rightharpoonup^\Gamma_\textsf{A}}}
\newcommand\DFtwoone[1]{{#1}^{\leftharpoonup^\tau_\textsf{A}}}
\newcommand\DFtwoonegamma[1]{{#1}^{\leftharpoonup^\Gamma_\textsf{A}}}
\newcommand\DFtwofun[1]{{#1}^{\rightharpoonup^{\tau}_{\textsf{B}}}}
\newcommand\DFtwofungamma[1]{{#1}^{\rightharpoonup^{\Gamma}_{\textsf{B}}}}
\newcommand\DFtwofunmeta[1]{{#1}^{\rightharpoonup^{\sigma}_\textsf{B}}}
\newcommand\DFfuntwo[1]{{#1}^{\leftharpoonup^{\tau}_\textsf{B}}}
\newcommand\DFfuntwogamma[1]{{#1}^{\leftharpoonup^{\Gamma}_\textsf{B}}}

\newcommand\DFfuntwometa[1]{{#1}^{\leftharpoonup^{\sigma}_\textsf{B}}}
\newcommand\fDtofunsuper[1]{\leftharpoonup^{#1}_\textsf{C}}
\newcommand\fDtofun[1]{({#1},\,{\gamma})^{\fDtofunsuper{\tau}}}
\newcommand\fDtofungamma[1]{{#1}^{\fDtofunsuper{\Gamma}}}
\newcommand\fDtofuntrail[1]{({#1},\,{\gamma})^{\fDtofunsuper{\mu}}}
\newcommand\fDtofunmeta[1]{({#1},\,{\gamma})^{\fDtofunsuper{\sigma}}}

\newcommand\funtofDsuper[1]{\rightharpoonup^{#1}_\textsf{C}}
\newcommand\funtofD[1]{({#1},\,{\mu},\,{\sigma})^{\funtofDsuper{\tau}}}

\newcommand\CPonetwosuper[1]{\rightharpoonup^{#1}_{\textsf{A$'$}}}
\newcommand\CPonetwo[1]{({#1},\,\gamma)^{\CPonetwosuper{\tau}}}
\newcommand\CPonetwotrail[1]{({#1},\,\gamma)^{\CPonetwosuper{\mu}}}

\newcommand\CPonetwogamma[1]{{#1}^{\CPonetwosuper{\Gamma}}}

\newcommand\idk{k_{id}}

\newcommand\Abs[2]{\lambda{#1}.\,{#2}}

\newcommand\ShiftX{\mathcal{S}}
\newcommand\Shift[2]{\mathcal{S}{#1}.\,{#2}}

\newcommand\ShiftZX{\mathcal{S}_{0}}
\newcommand\ShiftZ[2]{\mathcal{S}_{0}{#1}.\,{#2}}

\newcommand\ControlX{\mathcal{F}}
\newcommand\Control[2]{\mathcal{F}{#1}.\,{#2}}
\newcommand\ControlZX{\mathcal{F}_{0}}
\newcommand\ControlZ[2]{\mathcal{F}_{0}{#1}.\,{#2}}
\newcommand\myReset[1]{\langle{#1}\rangle}

\newcommand\Prompt[1]{\langle{#1}\rangle}

\renewcommand\Mc[6]{(({#1}\to\langle{#2},\,{#3}\rangle\,{#4})\times{#5})::{#6}}
\newcommand\Trail[4]{{#1}\to({#2},\,{#3})\,{#4}}
\newcommand\TrailMC[5]{{#1}\,\langle{#2}\rangle\,{#3}\,\langle{#4}\rangle\,{#5}} 

\newcommand\IdContType[4]{\textsf{id-cont-type}({#1},\,{#2},\,{#3},\,{#4})}
\newcommand\Compatible[3]{\textsf{compatible}({#1},\,{#2},\,{#3})}

\newcommand\Rule[1]{(\textsc{#1})}

\newcommand\JudgeSTLC[3]{{#1}\vdash{#2}:{#3}}
\newcommand\JudgeSTLCfDfun[3]{{#1}\vdash_{\textsf{4Dfun}}{#2}:{#3}}

\newcommand\ECPS[1]{\mathcal{E}[\![{#1}]\!]}
\newcommand\ECPSDash[1]{\mathcal{E'}[\![{#1}]\!]}
\newcommand\EfDfunCPS[1]{\mathcal{E}_{\textsf{4Dfun}}[\![{#1}]\!]}
\newcommand\EDFCPS[1]{\mathcal{E}_{\textsf{DF}}[\![{#1}]\!]}
\newcommand\EDFfunCPS[1]{\mathcal{E}_{\textsf{DF2}}[\![{#1}]\!]}



%% file: math.tex
\newcommand{\lambdaDSyntax}{
$
\begin{array}{rlcl}
    \textrm{Value} & v & := & n \ |\  x \ |\  \Abs{x}{e} \\[1ex]
    \textrm{Term} & e & := & v \ |\ e_1\,e_2 \ |\  \Shift{k}{e} \ |\  \Control{k}{e} \ |\  \ShiftZ{k}{e} \ |\  \ControlZ{k}{e} \ |\  \Prompt{e}\\[1ex]
    \textrm{Value and Term types} & \tau,\,\alpha,\,\beta & := & \mathbb{N} \ |\  \TypTrailMc{\tau_1\to\tau_2}{\mu_{\alpha}}{\sigma_{\alpha}}{\alpha}{\mu_{\beta}}{\sigma_{\beta}}{\beta}\\[1ex]
  \textrm{Trail types} & \mu & := & \bullet_{\mu} \ |\ \Arrow{\tau_1}{\mu}{\sigma}{\tau_2} \\[1ex]
  \textrm{Meta continuation types} & \sigma & := & \bullet_{\sigma}\ |\  \Mc{\tau_1}{\mu_1}{\sigma_1}{\tau_2}{\mu_2}{\sigma_2} \\[1ex]
\end{array}
$
}

\newcommand{\lambdaCSyntax}{
$
  \begin{array}{rrcl}
    \textrm{Value} & v & := & n \ | \  () \ |\  x \ |\  \Abs{x}{e} \\[1ex]
    \textrm{Terms} & e & := & v \ |\  e_1\,e_2 \ |\  (e_1,\,e_2) \ | \\[1ex]
    & & & (\CaseOf{e_1}{()}{e_2}{e_3}{e_4}) \\[1ex]
    \textrm{Types} & \tau & := & \texttt{int} \ |\  \texttt{unit} \ |\  \tau_1\to\tau_2 \ |\  \tau_1\times\tau_2 \\[1ex]
  \end{array}
$
}

\newcommand{\lambdaDSyntaxSR}{
$
\begin{array}{rccl}
    \textrm{Value} & v & := & n \ |\  x \ |\  \Abs{x}{e} \\[1ex]
    \textrm{Term} & e & := & v \ |\ e_1\,e_2 \ |\  \Shift{k}{e} \ |\  \Prompt{e}\\[1ex]
    \textrm{Value and Term types} & \tau & := & \mathbb{N} \ |\  \TypTrailMc{\tau_1\to\tau_2}{\mu_{\alpha}}{\sigma_{\alpha}}{\alpha}{\mu_{\beta}}{\sigma_{\beta}}{\beta}\\[1ex]
  \textrm{Trail types} & \mu & := & \bullet_{\mu} \ |\ \Arrow{\tau_1}{\mu}{\sigma}{\tau_2} \\[1ex]
  \textrm{Meta continuation types} & \sigma & := & \bullet_{\sigma}\ |\  \Mc{\tau_1}{\mu_1}{\sigma_1}{\tau_2}{\mu_2}{\sigma_2} \\[1ex]
\end{array}
$
}

\newcommand{\lambdaDSyntaxSRfun}{
$
\begin{array}{rccl}
    \textrm{Value} & v & := & n \ |\  x \ |\  \Abs{x}{e} \\[1ex]
    \textrm{Term} & e & := & v \ |\ e_1\,e_2 \ |\  \Shift{k}{e} \ |\  \Prompt{e}\\[1ex]
    \textrm{Value and Term types} & \tau & := & \mathbb{N} \ |\  \TypTrailMc{\tau_1\to\tau_2}{\mu_{\alpha}}{\sigma_{\alpha}}{\alpha}{\mu_{\beta}}{\sigma_{\beta}}{\beta}\\[1ex]
  \textrm{Trail types} & \mu & := & \bullet_{\mu} \ |\ \Arrow{\tau_1}{\mu}{\sigma}{\tau_2} \\[1ex]
  \textrm{Meta continuation types} & \sigma & := & \bullet_{\sigma}\ |\  \tau_1\to\tau_2\\[1ex]
\end{array}
$
}

\newcommand{\CpsTranslation}{
$
\begin{array}{rcl}
  \mathbb{N}^* & = & \text{\texttt{int}} \\
  (\TypTrailMc{\tau_1\to\tau_2}{\mu_{\alpha}}{\sigma_{\alpha}}{\alpha}{\mu_{\beta}}{\sigma_{\beta}}{\beta})^* &
  = & \tau_1^* \to (\tau_2^* \to \mu_{\alpha}^* \to \sigma_{\alpha}^* \to \alpha^*) \to \mu_{\beta}^* \to \sigma_{\beta}^* \to \beta^* \\[0.5em]
  \bullet_{\mu}^* & = & \texttt{unit} \\
  (\Arrow{\tau_1}{\mu}{\sigma}{\tau_2})^* & = & \tau_1^* \to \mu^* \to \sigma^* \to {\tau_2}^* \\[0.5em]
  \bullet_{\sigma}^* & = & \texttt{unit} \\
  (\Mc{\tau_1}{\mu_1}{\sigma_1}{\tau_2}{\mu_2}{\sigma_2})^* & = & ((\tau_1^* \to \mu_1^* \to \sigma_1^* \to {\tau_2}^*) \times {\mu_2}^*) \times {\sigma_2}^*
\end{array}
$
}

\newcommand{\CpsInterpreterColor}{
$
  \begin{array}{rcl}
    \idk & = & \Abs{v}{\Abs{t}{\Abs{m}{}}} \CaseOf{(t,\,m)}{((),\,())}{v}{((),\,((k_0,\, t_0),\, m_0)))}{k_0 \, v \, t_0 \, m_0 } \,|\,(t,\,m)\Rightarrow t \, v \, () \, m \\[1ex]
    \_@\_ & = & \Abs{t_1}{\Abs{t_2}{}} \CaseOf{(t_1,\,t_2)}{((),\,t_2)}{t_2}{(k,\,t_2)}{k::t_2} \\[1ex]
    \_\!::\!\_ & = & \Abs{k}{\Abs{t}{}} \CaseOf{(k,\,t)}{(k,\,())}{k}{(k,\,t)}{\Abs{v}{\Abs{t'}{\Abs{m'}{k \, v \, (t :: t')\, m'}}}}\\[1ex] 
  \end{array}
$
\\[2mm]
$
  \begin{array}{rcl}
    \ECPS{x}\,\rho\,\kappa &=& \Abs{t}{\Abs{m}{\kappa\,\rho(x)\,t\,m}}\\
    \ECPS{n}\,\rho\,\kappa &=& \Abs{t}{\Abs{m}{\kappa\,n\,t\,m}}\\

    \ECPS{\Abs{x}{e}}\,\rho\,\kappa &=& \Abs{t}{\Abs{m}{\kappa\,(\Abs{v}{\ECPS{e}\,\rho[v/x]})\,t\,m}}\\

    \ECPS{e_1\,e_2}\,\rho\,\kappa &=& \Abs{t}{\Abs{m}{}}
    \ECPS{e_1}\,\rho\,(\Abs{v_1}{\Abs{t_1}{\Abs{m_1}{}}}
    \ECPS{e_2}\,\rho\,(\Abs{v_2}{\Abs{t_2}{\Abs{m_2}{v_1\,v_2\,\kappa\,t_2\,m_2}}}) \,t_1\,m_1)\,t\,m\\

    \ECPS{\Shift{k}{e}} \,\rho\,\kappa  &=& \Abs{t}{\Abs{m}{}}
    \ECPS{e}\,\rho[\Abs{v}{\Abs{\kappa'}{\Abs{t'}{\Abs{m'}{\kappa \,v\,t\,\Bmath{\uwave{((\kappa',\,t') ,\, m')}} / k}}}}] \,\idk\,() \,m\\

    \ECPS{\Control{k}{e}} \,\rho\,\kappa  &=& \Abs{t}{\Abs{m}{}}
    \ECPS{e}\,\rho[\Abs{v}{\Abs{\kappa'}{\Abs{t'}{\Abs{m'}{\kappa \,v\,\Rmath{\underline{(t\,@\,(\kappa' :: t'))}}\, m' / k}}}}] \,\idk\,() \,m\\

    \ECPS{\ShiftZ{k}{e}} \,\rho\,\kappa  &=& \Abs{t}{\Abs{\Bmath{\uwave{((\kappa_0,\,t_0),\,m_0)}}}{}}
    \ECPS{e}\,\rho[\Abs{v}{\Abs{\kappa'}{\Abs{t'}{\Abs{m'}{\kappa \,v\,t\,\Bmath{\uwave{((\kappa',\,t') ,\, m')}} / k}}}}] \,\Bmath{\uwave{\,\kappa_0 \,t_0 \,m_0}}\\

    \ECPS{\ControlZ{k}{e}} \,\rho\,\kappa  &=& \Abs{t}{\Abs{\Bmath{\uwave{((\kappa_0,\,t_0),\,m_0)}}}{}}
    \ECPS{e}\,\rho[\Abs{v}{\Abs{\kappa'}{\Abs{t'}{\Abs{m'}{\kappa \,v\,\Rmath{\underline{(t\,@\,(\kappa' :: t'))}}\,m' / k}}}}] \,\Bmath{\uwave{\,\kappa_0 \,t_0 \,m_0}}\\

    \ECPS{\myReset{e}} \,\rho\,\kappa  &=&  \Abs{t}{\Abs{m}{}}
    \ECPS{e}\,\rho\,\,\idk\,()\,((\kappa,\,t)\,,\, m)
  \end{array}
$
}

\newcommand{\CPSInterpreterFDShift}{
  \[
  \begin{array}{l}
    \ECPS{\Shift{k}{e}} \,\rho\,\kappa \,=\,\Abs{t}{}\Abs{m}{}\\
    \quad
    \ECPS{e}\,\rho[\Abs{v}{}\Abs{\kappa'}{}\Abs{t'}{}\Abs{m'}{\kappa \,v\,t\,((\kappa',\,t'),\, m')} / k] \,k_{id}\,()\,m
  \\
  \quad\textit{where}
  \ \idk = \Abs{v}{\Abs{t}{\Abs{m}{}}} \texttt{case}\ {(t,\,m)}\ \texttt{of}\ ((),\,()) \Rightarrow v\\
  \quad\quad
  | \,((),\,((k_0,\, t_0),\, m_0))) \Rightarrow k_0 \, v \, t_0 \, m_0 \,
  | \,(t,\,m)\Rightarrow t \, v \, () \, m
  \end{array}
  \]
}

\newcommand{\CpsInterpreterDF}{
  \[
  \begin{array}{l}
    \EDFCPS{\Shift{k}{e}}\,\rho\,\kappa\,=\,
    \EDFCPS{e}\,\rho\,[
        \Abs{v}{}\Abs{\kappa'}{}
        \kappa'\,(\kappa\, v)/k
    ]\,(\Abs{v'}{v'})
  \end{array}
  \]
}

\newcommand{\CpsInterpreterDFFunc}{
  \[
  \begin{array}{l}
    \EDFfunCPS{\Shift{k}{e}}\,\rho\,\kappa\,=\,
    \Abs{m}{}\,\\
    \quad 
    \EDFfunCPS{e}\,\rho\,[
        \Abs{v}{}\Abs{\kappa'}{}\Abs{m'}{}
        \kappa\,v\,(\Abs{v'}{\kappa'\,v'\,m'})/k
    ]\quad\quad\\
    \hfill(\Abs{v''}{}\Abs{m''}{m''\,v''})\,m
  \end{array}
\]
}

\newcommand{\CpsInterpreterFDFunc}{
  \[
  \begin{array}{l}
    \EfDfunCPS{\Shift{k}{e}}\,\rho\,\kappa\,=\,
    \Abs{t}{}\Abs{m}{}\,\\
    \quad 
    \EfDfunCPS{e}\,\rho\,[
        \Abs{v}{}\Abs{\kappa'}{}\Abs{t'}{}\Abs{m'}{}
        \kappa\,v\,t\,(\Abs{v'}{\kappa'\,v'\,t'\,m'})/k
    ]\\
    \hfill k''_{id}\,()\,m
    \\
    \quad\textit{where}
    \ \idk = \Abs{v}{\Abs{t}{\Abs{m}{}}} \texttt{case}\ {(t,\,m)}\ \texttt{of}\ ((),\,()) \Rightarrow v\\
    \quad\quad
    | \,((),\,((k_0,\, t_0),\, m_0))) \Rightarrow k_0 \, v \, t_0 \, m_0 \,
    | \,(t,\,m)\Rightarrow t \, v \, () \, m
  \end{array}
  \]
}

\newcommand{\LambdaSZSyntaxOneColumn}{
  \[
    \begin{array}{rrcl}
      \text{Value} & v & := & n \ |\  x \ |\  \Abs{x}{e}\\
      \text{Term} & e & := & v \ |\  e_1\,e_2 \ |\  \ShiftZ{k}{e} \ |\  \myReset{e}\\
      \text{Type} & \tau & := & \mathbb{N} \ |\ \ArrowMB{\tau}{\sigma}{\tau'} \\
      \text{Type Annotation} & \sigma & := & \epsilon \ |\  \AnnMB{\tau}{\sigma}{\tau'}{\sigma'}
    \end{array}
  \]
}

\newcommand{\TypeSzRzMBOriginal}{
$
    \begin{array}{c}
      \infer[\Rule{MB-Var}]{ 
        \JudgeSTLC{\Gamma, \, x: \tau_{x}}
          {x}
          {\tau_{x}\AnnMB{\tau}{\sigma}{\tau}{\sigma}}
      }{} 
      \quad
      \infer[\Rule{MB-Abs}]{ 
        \JudgeSTLC{\Gamma}{\Abs{x}{e}}
          {\ArrowMB{\tau_1}{\sigma}{\tau_2} \AnnMB{\tau}{\sigma'}{\tau}{\sigma'}}
      }{ 
        \JudgeSTLC{\Gamma,\,x:\tau_1}{e}
          {\tau_2\,\sigma}
      }
      \\[1ex]
      \infer[\Rule{MB-App}]{ 
        \JudgeSTLC{\Gamma}{e_1\,e_2}{\tau_2\AnnMB{\tau'_1}{\sigma_1}{\tau'_2}{\sigma_2}}
      }{ 
        \JudgeSTLC
          {\Gamma}{e_1}
          {\ArrowMB{\tau_1}{\AnnMB{\tau'_1}{\sigma_1}{\tau'_3}{\sigma_3}}{\tau_2} \AnnMB{\tau'_4}{\sigma_4}{\tau'_2}{\sigma_2}}
        \quad
        \JudgeSTLC{\Gamma}{e_2}
          {\tau_1\AnnMB{\tau'_3}{\sigma_3}{\tau'_4}{\sigma_4}}
      }
      \\[1ex]
      \infer[\Rule{MB-Shift0}]{ 
        \JudgeSTLC{\Gamma}{\ShiftZ{k}{e}}{\tau_1\AnnMB{\tau_2}{\sigma_1}{\tau_3}{\sigma_2}}
      }{ 
        \JudgeSTLC{\Gamma,\,k:\ArrowMB{\tau_1}{\sigma_1}{\tau_2}}{e}{\tau_3\,\sigma_2}
      }
      \quad
      \infer[\Rule{MB-Reset0}]{ 
        \JudgeSTLC{\Gamma}{\myReset{e}}{\tau\,\sigma}
      }{ 
        \JudgeSTLC{\Gamma}{e}{\tau'\AnnMB{\tau'}{\AnnMB{\tau''}{\sigma''}{\tau''}{\sigma''}}{\tau}{\sigma}}
      }
    \end{array}
$
}

\newcommand{\TypeSzRzMBExtAbsSz}{
$
    \begin{array}{c}
      \infer[\Rule{MB-Abs-Ext}]{ 
        \JudgeSTLC{\Gamma}{\Abs{x}{e}}
          {\ArrowMB{\tau_1}{\AnnMB{\tau_3}{\sigma_3}{\tau_4}{\sigma_4}}{\tau_2} \AnnMB{\tau}{\sigma'}{\tau}{\sigma'}}
      }{ 
        \JudgeSTLC{\Gamma,\,x:\tau_1}{e}
          {\tau_2\,\AnnMB{\tau_3}{\sigma_3}{\tau_4}{\sigma_4}}
      }
      \\[1ex]
      \infer[\Rule{MB-Shift0-Ext}]{ 
        \JudgeSTLC{\Gamma}
          {\ShiftZ{k}{e}}
          {
            \tau_1
            \AnnMB
              {\tau_2}
              {\AnnMB{\tau_4}{\sigma_4}{\tau_5}{\sigma_5}}
              {\tau_3}
              {\AnnMB{\tau_6}{\sigma_6}{\tau_7}{\sigma_7}}
          }
      }{ 
        \JudgeSTLC
          {\Gamma,\,k:\ArrowMB{\tau_1}{\AnnMB{\tau_4}{\sigma_4}{\tau_5}{\sigma_5}}{\tau_2}}
          {e}
          {\tau_3\,\AnnMB{\tau_6}{\sigma_6}{\tau_7}{\sigma_7}}
      }
    \end{array}
$
}

\newcommand{\TypeSzRzMBTransToMB}{
$
    \begin{array}{rcl}
      \toMBT{\mathbb{N}} & = & \mathbb{N}
      \\
      \toMBT{(\TrailMC{\tau_1\to\tau_2}{\sigma_{\alpha}}{\alpha}{\sigma_{\beta}}{\beta})}
      & = &
      \ArrowMB
        {\toMBT{\tau_1}}
        {\AnnMB{\tau_3}{\sigma_3}{\tau_4}{\sigma_4}}
        {\toMBT{\tau_2}}  
      \quad
      \\
      & & 
      \textit{where} \quad 
      (\tau_3,\,\sigma_3) = \toMBA{\sigma_{\alpha}}{\alpha}, \quad
      (\tau_4,\,\sigma_4) = \toMBA{\sigma_{\beta}}{\beta}
      \\[1ex]
      \toMBA{\bullet_{\sigma}}{\tau} & = & (\toMBT{\tau},\,\epsilon)
      \\
      \toMBA{((\tau_1\to\langle\sigma_1\rangle\,\tau_2)::\sigma_2)}{\tau}
      & = &
      (\toMBT{\tau_1},\,\AnnMB{\tau_3}{\sigma_3}{\tau_4}{\sigma_4})
      \\
      & &
      \textit{where} \quad (\tau_3,\,\sigma_3) = \toMBA{\sigma_1}{\tau_2}, \quad 
      (\tau_4,\,\sigma_4) = \toMBA{\sigma_2}{\tau}
      \\
      \toMBG{\bullet_{\Gamma}} & = & \bullet_{\Gamma}\\
      \toMBG{(x\,:\,\tau,\,\Gamma)} & = & x\,:\,\toMBT{\tau},\,\toMBG{\Gamma}\\
    \end{array}
$
}

\newcommand{\TypeSzRzMBTransFromMB}{
$
    \begin{array}{rcl}
      \toFDT{\mathbb{N}} & = & \mathbb{N} 
      \\
      \toFDT{(\ArrowMB{\tau_1}{\AnnMB{\tau_3}{\sigma_3}{\tau_4}{\sigma_4}}{\tau_2})}
      & = &
      \TrailMC
        {(\toFDT{\tau_1})\to(\toFDT{\tau_2})}
        {\sigma_{\alpha}}{\alpha}{\sigma_{\beta}}{\beta}
      \\
      &  & 
      \textit{where} \quad 
      (\sigma_{\alpha},\,\alpha) = \toFDA{\tau_3}{\sigma_3}, \quad
      (\sigma_{\beta},\,\beta) = \toFDA{\tau_4}{\sigma_4}
      \\[1ex]
      \toFDA{\tau}{\epsilon} & = & (\bullet_{\sigma},\,\toFDT{\tau})
      \\
      \toFDA{\tau}{(\AnnMB{\tau_1}{\sigma_1}{\tau_2}{\sigma_2})} 
      & = &
      (((\toFDT{\tau}\,\to\langle\sigma_3\rangle\,\tau_3)::\sigma_4),\,\tau_4)
      \\
      & &
      \textit{where} \quad 
      (\sigma_3,\,\tau_3) = \toFDA{\tau_1}{\sigma_1}, \quad
      (\sigma_4,\,\tau_4) = \toFDA{\tau_2}{\sigma_2}\\
      \toFDG{\bullet_{\Gamma}} & = & \bullet_{\Gamma}\\
      \toFDG{(x\,:\,\tau,\,\Gamma)} & = & x\,:\,\toFDT{\tau},\,\toFDG{\Gamma}\\
    \end{array}
$
}

\newcommand{\TypeSzRzFromfDdash}{
$
\begin{array}{rcl}
    \fDdashfD{\mathbb{N}}  &=& \mathbb{N}\\
    \fDdashfD{(\TrailMC{\tau_2\to\tau_1}{\sigma_1}{\alpha}{\sigma_2}{\beta})} &=& 
    \TypTrailMc{\fDdashfD{\tau_2}\to\fDdashfD{\tau_1}}{\bullet_{\mu}}{\fDdashfDmeta{\sigma_1}}{\fDdashfD{\alpha}}{\bullet_{\mu}}{\fDdashfDmeta{\sigma_2}}{\fDdashfD{\beta}}\\[1ex]
    \fDdashfDmeta{\bullet_{\sigma}} &=& \bullet_{\sigma}\\
    \fDdashfDmeta{(\tau_1\to\langle\sigma_1\rangle\,\tau_2::\sigma_2)} &=& 
    \Mc{\fDdashfD{\tau_1}}{\bullet_{\mu}}{\fDdashfDmeta{\sigma_1}}{\fDdashfD{\tau_2}}{\bullet_{\mu}}{\fDdashfDmeta{\sigma_2}}\\[1ex]
    \fDdashfDgamma{\bullet_{\Gamma}} &=& \bullet_{\Gamma}\\
    \fDdashfDgamma{x\,:\,\tau,\,\Gamma} &=& x\,:\,\fDdashfD{\tau},\,\fDdashfDgamma{\Gamma}\\[1em]
    \fDfDdash{\mathbb{N}} &=& \mathbb{N}\\
    \fDfDdash{(\TypTrailMc{\tau_1\to\tau_2}{\mu_1}{\sigma_1}{\alpha}{\mu_2}{\sigma_2}{\beta})} &=&
    \TrailMC{\fDfDdash{\tau_1}\to\fDfDdash{\tau_2}}{\fDfDdashmeta{\sigma_1}}{\fDfDdash{\alpha}}{\fDfDdashmeta{\sigma_2}}{\fDfDdash{\beta}}\\[1ex]
    \fDfDdashmeta{\bullet_{\sigma}} &=& \bullet_{\sigma} \\
    \fDfDdashmeta{(\Mc{\tau_1}{\mu_1}{\sigma_1}{\tau_2}{\mu_2}{\sigma_2})} &=&
    \fDfDdash{\tau_1}\to\langle\fDfDdashmeta{\sigma_1}\rangle\,\fDfDdash{\tau_2}::\fDfDdashmeta{\sigma_2}\\[1ex]
    \fDfDdashgamma{\bullet_{\Gamma}} &=& \bullet_{\Gamma}\\
    \fDfDdashgamma{x\,:\,\tau,\,\Gamma} &=& x\,:\,\fDfDdash{\tau},\,\fDfDdashgamma{\Gamma}
\end{array}
$
}

\newcommand{\TypeSRDFTrans}{
$
\begin{array}{rcl}
    \DFonetwo{\mathbb{N}}{\gamma} & = & \mathbb{N} \\
    \DFonetwo{((\tau_1\to\tau_2),\,\alpha,\,\beta)}{\gamma}
    & = &
    \TrailMC{\DFonetwo{\tau_1}{\gamma}\to\DFonetwo{\tau_2}{\gamma}}
        {\DFonetwo{\alpha}{\gamma}\to\Rtext{\gamma}}
        {\Rtext{\gamma}}
        {\DFonetwo{\beta}{\gamma}\to\Rtext{\gamma}}
        {\Rtext{\gamma}}\\[1ex]
    \DFonetwogamma{\bullet_{\Gamma}} &=& \bullet_{\Gamma}\\
    \DFonetwogamma{(x\,:\,\tau,\,\Gamma)} &=& x\,:\,\DFonetwo{\tau}{\gamma},\,\DFonetwogamma{\Gamma}\\[1em]
    \DFtwoone{\mathbb{N}} & = & \mathbb{N}\\
    \DFtwoone{\TrailMC{\tau_1\to\tau_2}{\alpha\to\gamma_1}{\gamma_2}{\beta\to\gamma_3}{\gamma_4}} &=& 
    (\DFtwoone{\tau_1}\to\DFtwoone{\tau_2}),\,\DFtwoone{\alpha},\,\DFtwoone{\beta}
    \\[1ex]
    \DFtwoonegamma{\bullet_{\Gamma}} &=& \bullet_{\Gamma}\\
    \DFtwoonegamma{(x\,:\,\tau,\,\Gamma)} &=& x\,:\,\DFtwoone{\tau},\,\DFtwoonegamma{\Gamma}
\end{array}
$
}

\newcommand{\TypeSRTrailTrans}{
$
\begin{array}{rcl}
    \DFtwofun{\mathbb{N}} & = & \mathbb{N}\\
    \DFtwofun{(\TrailMC{\tau_1\to\tau_2}{\sigma_1}{\alpha}{\sigma_2}{\beta})} & = & \TypTrailMc{\DFtwofun{\tau_1}\to\DFtwofun{\tau_2}}{\bullet_{\mu}}{\DFtwofunmeta{\sigma_1}}{\DFtwofun{\alpha}}{\bullet_{\mu}}{\DFtwofunmeta{\sigma_2}}{\DFtwofun{\beta}}\\[1ex]
    \DFtwofunmeta{(\tau_1\to\tau_2)} &=&
    \DFtwofun{\tau_1} \to \DFtwofun{\tau_2}\\[1ex]
    \DFtwofungamma{\bullet_{\Gamma}} &=& \bullet_{\Gamma}\\
    \DFtwofungamma{(x\,:\,\tau,\,\Gamma)} &=& x\,:\,\DFtwofun{\tau},\,\DFtwofungamma{\Gamma}\\[1em]
    \DFfuntwo{\mathbb{N}} & = & \mathbb{N}\\
    \DFfuntwo{(\TypTrailMc{\tau_1\to\tau_2}{\mu_1}{\sigma_1}{\alpha}{\mu_2}{\sigma_2}{\beta})} & = & \TrailMC{\DFfuntwo{\tau_1}\to\DFfuntwo{\tau_2}}{\DFfuntwometa{\sigma_1}}{\DFfuntwo{\alpha}}{\DFfuntwometa{\sigma_2}}{\DFfuntwo{\beta}}\\[1ex]
    \DFfuntwometa{(\tau_1\to\tau_2)} &=&
    \DFfuntwo{\tau_1} \to \DFfuntwo{\tau_2} \\[1ex]
    \DFfuntwogamma{\bullet_{\Gamma}} &=& \bullet_{\Gamma}\\
    \DFfuntwogamma{(x\,:\,\tau,\,\Gamma)} &=& x\,:\,\DFfuntwo{\tau},\,\DFfuntwogamma{\Gamma}
    \end{array}
$
}

\newcommand{\TypeSRTransTofun}{
$
\begin{array}{rcl}
    \fDtofun{\mathbb{N}} & = & \mathbb{N}\\
    \fDtofun{\TypTrailMc{\tau_1\to\tau_2}{\mu_1}{\sigma_1}{\alpha}{\mu_2}{\sigma_2}{\beta}} & = & 
    \fDtofun{\tau_1}\to\fDtofun{\tau_2}
    \langle\fDtofuntrail{\mu_1},\,\fDtofunmeta{\sigma_1}\rangle\,\fDtofun{\alpha}
    \\ & & \hfill 
    \langle\fDtofuntrail{\mu_2},\,\fDtofunmeta{\sigma_2}\rangle\,\fDtofun{\beta}
    \\[1ex]
    \fDtofuntrail{\bullet_{\mu}} & = & \bullet_{\mu} \\
    \fDtofuntrail{\Arrow{\tau_1}{\mu_1}{\sigma_1}{\tau_2}} & = & \Arrow{\fDtofun{\tau_1}}{\fDtofuntrail{\mu_1}}{\fDtofunmeta{\sigma_1}}{\fDtofun{\tau_2}} \\[1ex]
    \fDtofunmeta{\bullet_{\sigma}}& = & \Rtext{\gamma \to \gamma} \\
    \fDtofunmeta{\Mc{\tau_1}{\mu_1}{\sigma_1}{\tau_2}{\mu_2}{\sigma_2}} &=&
    \fDtofunmeta{\tau_1} \to \fDtofun{\tau_2}
    \\[1ex]
    \fDtofungamma{\bullet_{\Gamma}} &=& \bullet_{\Gamma}\\
    \fDtofungamma{(x\,:\,\tau,\,\Gamma)} &=& x\,:\,\fDtofun{\tau},\,\fDtofungamma{\Gamma}
\end{array}
$
}

\newcommand{\TypeCPTransOneTwo}{
$
\begin{array}{rcl}
    \CPonetwo{\mathbb{N}} &=& \mathbb{N} \\
    \CPonetwo{(\TrailMC{\tau}{\mu_1}{\alpha}{\mu_2}{\beta})} &=& 
    \CPonetwo{\tau}\langle\CPonetwo{\mu_1},\,(\CPonetwo{\alpha}\to\Rtext{\gamma})\rangle\,\Rtext{\gamma}\,
    \langle\CPonetwo{\mu_2},\,(\CPonetwo{\beta}\to\Rtext{\gamma})\rangle\,\Rtext{\gamma}
    \\[1ex]
    \CPonetwotrail{\bullet_{\mu}} &=& \bullet_{\mu}\\
    \CPonetwotrail{(\tau_1\to\langle\,\mu\,\rangle\,\tau_2)} &=&
    \Arrow{\CPonetwo{\tau_1}}{(\CPonetwotrail{\mu}}{\CPonetwo{\tau_1}\to\gamma)}{\gamma}
    \\[1ex]
    \CPonetwogamma{\bullet_{\Gamma}} &=& \bullet_{\Gamma}\\
    \CPonetwogamma{(x\,:\,\tau,\,\Gamma)} &=& x\,:\,\CPonetwo{\tau},\,\CPonetwogamma{\Gamma}
\end{array}
$
}

\newcommand{\InferCompatible}{
  \[
  \begin{array}{l}
    k^{\Rmath{(\Arrow{\tau_1}{\mu_1}{\sigma_1}{\tau'_1})^*}} :: 
    t^{\Rmath{(\Arrow{\tau_2}{\mu_2}{\sigma_2}{\tau'_2})^*}}  
    \hfill\\
    =
    \Abs
      {v^{\Rmath{\tau_3^*}}}
      {\Abs
        {{t'}^{\Rmath{\mu_3^*}}}
        {\Abs
          {{m'}^{\Rmath{\sigma_3^*}}}{}}}
          ~\\
          \quad\quad
            (
              k^{\Rmath{(\Arrow{\tau_1}{\mu_1}{\sigma_1}{\tau'_1})^*}}\, 
              v^{\Rmath{\tau_3^*}} \,
              (
                t^{\Rmath{(\Arrow{\tau_2}{\mu_2}{\sigma_2}{\tau'_2})^*}} :: 
                {t'}^{\Rmath{\mu_3^*}}
              )\, 
              {m'}^{\Rmath{\sigma_3^*}}
            )^{\Rmath{{\tau'_3}^*}}
    \end{array}
  \]
}

\newcommand{\InferIdContType}{
  \[
    \begin{array}{l}
      (
        \idk \,\,\,
        v^{\Rmath{\tau^*}} \, 
        ()^{\Rmath{\bullet_{\mu}^*}} \, 
        ((k_0,\, t_0),\, m_0)^{\Rmath{(\Mc{\tau_1}{\mu_1}{\sigma_1}{\tau'_1}{\mu_2}{\sigma_2})^*}}
      )^{\Rmath{{\tau'}^*}} 
      \\ 
      =\,
        k_0^{\Rmath{(\Arrow{\tau_1}{\mu_1}{\sigma_1}{\tau'_1})^*}} \, 
        v^{\Rmath{\tau^*}} \, 
        t_0^{\Rmath{\mu_2^*}} \, 
        m_0^{\Rmath{\sigma_2^*}}
      \\
    \end{array}
  \]
}

\newcommand{\TypeLambdaVariable}{
$
  \begin{array}{c}
    \hspace{15em}
    \infer[\Rule{TVar}]{
        \JudgeSTLC
          {\Gamma}{x}
          {
            \TypTrailMc
              {\tau}
              {\mu_{\alpha}}{\sigma_{\alpha}}
              {\alpha}
              {\mu_{\alpha}}{\sigma_{\alpha}}
              {\alpha}
          }
      }{ 
        x \, : \, \tau \in \Gamma
      }
    \quad\quad
    \infer[\Rule{TNum}]{
      \JudgeSTLC
        {\Gamma}{n}
        {\TypTrailMc{\mathbb{N}}{\mu_{\alpha}}{\sigma_{\alpha}}{\alpha}{\mu_{\alpha}}{\sigma_{\alpha}}{\alpha}}}
      {} 
    \\[1em]
    \infer[\Rule{TLam}]{ 
      \JudgeSTLC{\Gamma}{(\lambda\,x.\,e)}{\TypTrailMc{(\TypTrailMc{\tau_1\to\tau_2}{\mu_{\alpha}}{\sigma_{\alpha}}{\alpha}{\mu_{\beta}}{\sigma_{\beta}}{\beta})}{\mu_{\gamma}}{\sigma_{\gamma}}{\gamma}{\mu_{\gamma}}{\sigma_{\gamma}}{\gamma}}
    }{ 
      \JudgeSTLC{\Gamma,\,x\,:\,\tau_1}{e}{\TypTrailMc{\tau_2}{\mu_{\alpha}}{\sigma_{\alpha}}{\alpha}{\mu_{\beta}}{\sigma_{\beta}}{\beta}}
    }
    \\[1em]
    \infer[\Rule{TApp}]{ 
      \begin{array}{c}
        \JudgeSTLC{\Gamma}{(e_1\,e_2)}{\TypTrailMc{\tau_2}{\mu_{\alpha}}{\sigma_{\alpha}}{\alpha}{\mu_{\delta}}{\sigma_{\delta}}{\delta}} 
      \end{array}
        }{ 
        \JudgeSTLC{\Gamma}{e_1}{\TypTrailMc{(\TypTrailMc{\tau_1\to\tau_2}{\mu_{\alpha}}{\sigma_{\alpha}}{\alpha}{\mu_{\beta}}{\sigma_{\beta}}{\beta})}{\mu_{\gamma}}{\sigma_{\gamma}}{\gamma}{\mu_{\delta}}{\sigma_{\delta}}{\delta}}
        \quad
        \JudgeSTLC{\Gamma}{e_2}{\TypTrailMc{\tau_1}{\mu_{\beta}}{\sigma_{\beta}}{\beta}{\mu_{\gamma}}{\sigma_{\gamma}}{\gamma}}
    }\\[0.5ex]
  \end{array}
$
}

\newcommand{\TypeLambdaVariablefDfun}{
$
  \begin{array}{c}
    \text{\fbox{$\Gamma\,\vdash_{\textsf{4Dfun}}\,e\,:\,\TypTrailMc{\tau}{\bullet_\mu}{\sigma_{\alpha}}{\alpha}{\bullet_\mu}{\sigma_{\beta}}{\beta}$}} \hspace{35em}\Wmath{a}\\[1em]
    \infer[\Rule{4Dfun-Var}]{
        \JudgeSTLCfDfun
          {\Gamma}{x}
          {
            \TypTrailMc
              {\tau}
              {\bullet_{\mu}}{\sigma_{\alpha}}
              {\alpha}
              {\bullet_{\mu}}{\sigma_{\alpha}}
              {\alpha}
          }
      }{ 
        x \, : \, \tau \in \Gamma
      }
    \quad\quad
    \infer[\Rule{4Dfun-Num}]{
      \JudgeSTLCfDfun
        {\Gamma}{n}
        {\TypTrailMc{\mathbb{N}}{\bullet_{\mu}}{\sigma_{\alpha}}{\alpha}{\bullet_{\mu}}{\sigma_{\alpha}}{\alpha}}}
      {} 
    \\[1em]
    \infer[\Rule{4Dfun-Lam}]{ 
      \JudgeSTLCfDfun{\Gamma}{(\lambda\,x.\,e)}{\TypTrailMc{(\TypTrailMc{\tau_1\to\tau_2}{\bullet_{\mu}}{\sigma_{\alpha}}{\alpha}{\bullet_{\mu}}{\sigma_{\beta}}{\beta})}{\bullet_{\mu}}{\sigma_{\gamma}}{\gamma}{\bullet_{\mu}}{\sigma_{\gamma}}{\gamma}}
    }{ 
      \JudgeSTLCfDfun{\Gamma,\,x\,:\,\tau_1}{e}{\TypTrailMc{\tau_2}{\bullet_{\mu}}{\sigma_{\alpha}}{\alpha}{\bullet_{\mu}}{\sigma_{\beta}}{\beta}}
    }
    \\[1em]
    \infer[\Rule{4Dfun-App}]{ 
      \begin{array}{c}
        \JudgeSTLCfDfun{\Gamma}{(e_1\,e_2)}{\TypTrailMc{\tau_2}{\bullet_{\mu}}{\sigma_{\alpha}}{\alpha}{\bullet_{\mu}}{\sigma_{\delta}}{\delta}} 
      \end{array}
        }{ 
        \JudgeSTLCfDfun{\Gamma}{e_1}{\TypTrailMc{(\TypTrailMc{\tau_1\to\tau_2}{\bullet_{\mu}}{\sigma_{\alpha}}{\alpha}{\bullet_{\mu}}{\sigma_{\beta}}{\beta})}{\bullet_{\mu}}{\sigma_{\gamma}}{\gamma}{\bullet_{\mu}}{\sigma_{\delta}}{\delta}}
        \quad
        \JudgeSTLCfDfun{\Gamma}{e_2}{\TypTrailMc{\tau_1}{\bullet_{\mu}}{\sigma_{\beta}}{\beta}{\bullet_{\mu}}{\sigma_{\gamma}}{\gamma}}
    }\\[0.5ex]
  \end{array}
$
}

\newcommand{\TypeLambdaVariablefDdash}{
$
  \begin{array}{c}
    \hspace{15em}
    \infer[\Rule{TVar$'$}]{
        \JudgeSTLC
          {\Gamma}{x}
          {
            \TypTrailMc
              {\tau}
              {\bullet_{\mu}}{\sigma_{\alpha}}
              {\alpha}
              {\bullet_{\mu}}{\sigma_{\alpha}}
              {\alpha}
          }
      }{ 
        x \, : \, \tau \in \Gamma
      }
    \quad\quad
    \infer[\Rule{TNum$'$}]{
      \JudgeSTLC
        {\Gamma}{n}
        {\TypTrailMc{\mathbb{N}}{\bullet_{\mu}}{\sigma_{\alpha}}{\alpha}{\bullet_{\mu}}{\sigma_{\alpha}}{\alpha}}}
      {} 
    \\[1em]
    \infer[\Rule{TLam$'$}]{ 
      \JudgeSTLC{\Gamma}{(\lambda\,x.\,e)}{\TypTrailMc{(\TypTrailMc{\tau_1\to\tau_2}{\bullet_{\mu}}{\sigma_{\alpha}}{\alpha}{\bullet_{\mu}}{\sigma_{\beta}}{\beta})}{\bullet_{\mu}}{\sigma_{\gamma}}{\gamma}{\bullet_{\mu}}{\sigma_{\gamma}}{\gamma}}
    }{ 
      \JudgeSTLC{\Gamma,\,x\,:\,\tau_1}{e}{\TypTrailMc{\tau_2}{\bullet_{\mu}}{\sigma_{\alpha}}{\alpha}{\bullet_{\mu}}{\sigma_{\beta}}{\beta}}
    }
    \\[1em]
    \infer[\Rule{TApp$'$}]{ 
      \begin{array}{c}
        \JudgeSTLC{\Gamma}{(e_1\,e_2)}{\TypTrailMc{\tau_2}{\bullet_{\mu}}{\sigma_{\alpha}}{\alpha}{\bullet_{\mu}}{\sigma_{\delta}}{\delta}} 
      \end{array}
        }{ 
        \JudgeSTLC{\Gamma}{e_1}{\TypTrailMc{(\TypTrailMc{\tau_1\to\tau_2}{\bullet_{\mu}}{\sigma_{\alpha}}{\alpha}{\bullet_{\mu}}{\sigma_{\beta}}{\beta})}{\bullet_{\mu}}{\sigma_{\gamma}}{\gamma}{\bullet_{\mu}}{\sigma_{\delta}}{\delta}}
        \quad
        \JudgeSTLC{\Gamma}{e_2}{\TypTrailMc{\tau_1}{\bullet_{\mu}}{\sigma_{\beta}}{\beta}{\bullet_{\mu}}{\sigma_{\gamma}}{\gamma}}
    }\\[0.5ex]
  \end{array}
$
}

\newcommand{\TypeSVariable}{
$
    \begin{array}{c}
    \infer[\Rule{TShift}]
    { 
        \JudgeSTLC
            {\Gamma}
            {\Shift{k}{e}}
            {
                \TypTrailMc{\tau}{\mu_{\beta}}
                    {(\Mc{\tau_1}{\mu_1}{\sigma_1}{\tau_2}{\mu_2}{\sigma_2})}
                    {\alpha}{\mu_{\beta}}{\sigma_{\beta}}{\beta}
            }
    }
    { 
      \begin{array}{c}
        \IdContType{\gamma}{\mu_{id}}{\sigma_{id}}{\gamma'}
        \\
        \JudgeSTLC
        {
          \Gamma ,\,k\,:\,
          \TypTrailMc
            {\tau \to \tau_1}{\mu_1}{\sigma_1}{\tau_2}{\mu_2}{\sigma_2}{\alpha}
        }
        {e}
        {
          \TypTrailMc
            {\gamma}
            {\mu_{id}}{\sigma_{id}}
            {\gamma'}
            {\bullet_{\mu}}{\sigma_{\beta}}
            {\beta}
        }
      \end{array}
    }\\[1ex]
    \end{array}
$
}

\newcommand{\TypeSVariablefDfun}{
$
    \begin{array}{c}
    \infer[\Rule{4Dfun-Shift}]
    { 
        \JudgeSTLCfDfun
            {\Gamma}
            {\Shift{k}{e}}
            {
                \TypTrailMc{\tau}{\bullet_{\mu}}
                    {(\tau_1\to\tau_2)}
                    {\alpha}{\bullet_{\mu}}{\sigma_{\beta}}{\beta}
            }
    }
    { 
      \begin{array}{c}
        \IdContType{\gamma}{\bullet_{\mu}}{\sigma_{id}}{\gamma'}
        \\
        \JudgeSTLCfDfun
        {
          \Gamma ,\,k\,:\,
          \TypTrailMc
            {\tau \to \tau_1}{\bullet_{\mu}}{\sigma_1}{\tau_2}{\bullet_{\mu}}{\sigma_1}{\alpha}
        }
        {e}
        {
          \TypTrailMc
            {\gamma}
            {\bullet_{\mu}}{\sigma_{id}}
            {\gamma'}
            {\bullet_{\mu}}{\sigma_{\beta}}
            {\beta}
        }
      \end{array}
    }\\[1ex]
    \end{array}
$
}

\newcommand{\TypeSVariablefDdash}{
$
    \begin{array}{c}
    \infer[\Rule{TShift$'$}]
    { 
        \JudgeSTLC
            {\Gamma}
            {\Shift{k}{e}}
            {
                \TypTrailMc{\tau}{\bullet_{\mu}}
                    {(\Mc{\tau_1}{\bullet_{\mu}}{\sigma_1}{\tau_2}{\bullet_{\mu}}{\sigma_1})}
                    {\alpha}{\bullet_{\mu}}{\sigma_{\beta}}{\beta}
            }
    }
    { 
      \begin{array}{c}
        \IdContType{\gamma}{\bullet_{\mu}}{\sigma_{id}}{\gamma'}
        \\
        \JudgeSTLC
        {
          \Gamma ,\,k\,:\,
          \TypTrailMc
            {\tau \to \tau_1}{\bullet_{\mu}}{\sigma_1}{\tau_2}{\bullet_{\mu}}{\sigma_1}{\alpha}
        }
        {e}
        {
          \TypTrailMc
            {\gamma}
            {\bullet_{\mu}}{\sigma_{id}}
            {\gamma'}
            {\bullet_{\mu}}{\sigma_{\beta}}
            {\beta}
        }
      \end{array}
    }\\[1ex]
    \end{array}
$
}

\newcommand{\TypeSDashVariable}{
\[
    \begin{array}{c}
    \infer[]
    { 
        \begin{array}{l}
        \Gamma\,\vdash\,
        \Shift{k}{e}\,:\,
        \\
        \quad\quad
        \TypTrailMc{\tau}{\mu_{\beta}}
            {({\tau_1}\to\langle{\uwave{\bullet_{\mu},\,\sigma_1}}\rangle\,{\tau_2})\,\times\,\uline{\bullet_{\mu}::\sigma_1}}
            {\alpha}{\mu_{\beta}}{\sigma_{\beta}}{\beta}
        \end{array}
    }
    { 
      \begin{array}{l}
        \IdContType{\gamma}{\mu_{id}}{\sigma_{id}}{\gamma'}
        \\
        \Gamma ,\,k\,:\,
        \TrailMC
        {\tau \to \tau_1}{\uwave{{\bullet_{\mu}},\,\sigma_1}}{\tau_1}{\uline{\bullet_{\mu}\,,\,\sigma_1}}{\alpha}
        \\[-1ex]
        \quad\quad\quad
        \vdash\,e\,:\,
        \TrailMC
            {\gamma}
            {\mu_{id},\,\sigma_{id}}
            {\gamma'}
            {\bullet_{\mu},\,\sigma_{\beta}}
            {\beta}
        \hspace{2em}\Rule{TShift$'$}
      \end{array}
    }
    \end{array}
\]
}

\newcommand{\TypeSDFWoTrailMc}{
\[
    \begin{array}{c}
    \infer[\Rule{DF-Shift}]
    { 
            \Gamma\,\vdash_{\textsf{DF}}\,
            \Shift{k}{e}\,:\,
              \tau,\,\alpha,\,\beta
    }
    { 
        \begin{array}{c}
            {      
                \Gamma ,\,k\,:\, \tau\to\alpha,\,\delta,\,\delta
            }\,
            \vdash_{\textsf{DF}}\,
            {e}\,:\,
            {
              \epsilon,\,\epsilon,\,\beta
            }
        \end{array}
    }
    \end{array}
\]
}

\newcommand{\TypeSDFFunc}{
\[
    \begin{array}{c}
    \infer[]
    {
        \begin{array}{l}
        \Gamma\vdash_{\textsf{DF2}}\,\Shift{k}{e}\,:\,
        \TrailMC{\tau}{\tau_1\to\tau_2}{\alpha}{\sigma_{\beta}}{\beta}
        \\[-1.3em]
        \hspace{17em}\Rule{DF2-Shift}
        \end{array}
    }
    {
        \Gamma,\,k\,:\,\TrailMC{\tau\to\tau_1}{\sigma_1}{\tau_2}{\sigma_1}{\alpha}\,
        \vdash_{\textsf{DF2}}\,e\,:\,\TrailMC{\gamma}{\gamma\to\gamma'}{\gamma'}{\sigma_{\beta}}{\beta}
    }
    \end{array}
\]
}

\newcommand{\TypeSfDWoTrailMc}{
\[
    \begin{array}{c}
    \infer[]
    {
        \begin{array}{l}
        \Gamma\vdash_{\textsf{4Dfun}}\,\Shift{k}{e}\,:\,
        \TypTrailMc{\tau}{\bullet_{\mu}}{\tau_1\to\tau_2}{\alpha}{\bullet_{\mu}}{\sigma_{\beta}}{\beta}
        \end{array}
    }
    {
        \begin{array}{l}
        \textsf{id-cont-type}(\gamma,\,\mu_{id},\,\sigma_{id},\,\gamma')
        \\
        \Gamma,\,k\,:\,\TypTrailMc{\tau\to\tau_1}{\bullet_{\mu}}{\sigma_1}{\tau_2}{\bullet_{\mu}}{\sigma_1}{\alpha}\,
        \\
        \quad
        \vdash_{\textsf{4Dfun}}\,e\,:\,\TypTrailMc{\gamma}{\bullet_{\mu}}{\sigma_{id}}{\gamma'}{\bullet_{\mu}}{\sigma_{\beta}}{\beta} 
        \\[-1em]
        \hspace{16em} \Rule{4Dfun-Shift}
        \end{array}
    }
    \end{array}
\]
}

\newcommand{\TypeCVariable}{
$
    \begin{array}{c}
    \infer[\Rule{TControl}]
    { 
        \JudgeSTLC
            {\Gamma}
            {\Control{k}{e}}
            {\TypTrailMc{\tau}{\mu_{\alpha}}{\sigma_{\alpha}}{\alpha}{\mu_{\beta}}{\sigma_{\beta}}{\beta}}
    }
    { 
      \begin{array}{c}
        \IdContType{\gamma}{\mu_{id}}{\sigma_{id}}{\gamma'}
        \\
        \Compatible{(\Arrow{\tau_1}{\mu_1}{\sigma_1}{\tau_2})}{\mu_2}{\mu_{\gamma}}
        \quad
        \Compatible{\mu_{\beta}}{\mu_{\gamma}}{\mu_{\alpha}}
        \\
        \JudgeSTLC
            {
                \Gamma ,\,k\,:\,
                \TypTrailMc
                    {\tau \to \tau_1}{\mu_1}{\sigma_1}{\tau_2}{\mu_2}{\sigma_{\alpha}}{\alpha}
            }
            {e}
            {
                 \TypTrailMc{\gamma}{\mu_{id}}{\sigma_{id}}{\gamma'}{\bullet_{\mu}}{\sigma_{\beta}}{\beta}
            }
      \end{array}
    }
    \\[1ex]
    \end{array}
$
}

\newcommand{\TypeSzVariable}{
$
    \begin{array}{c}
    \infer[\Rule{TShift0}]
        { 
            \JudgeSTLC
                {\Gamma}
                {\ShiftZ{k}{e}} 
                {\TypTrailMc
                        {\tau}
                        {\mu_{\beta}}
                        {(\Mc{\tau_1}{\mu_1}{\sigma_1}{\tau_2}{\mu_2}{\sigma_2})}
                        {\alpha}
                        {\mu_{\beta}}
                        {(\Mc{\tau_0}{\mu_0}{\sigma_0}{\tau'_0}{\mu'_0}{\sigma'_0})}
                        {\beta}
                }
        }
        { 
          \begin{array}{l}
            \hspace{2em}
            \JudgeSTLC
                {
                    \Gamma ,\,k\,:\,
                    \TypTrailMc{\tau\to\tau_1}{\mu_1}{\sigma_1}{\tau_2}{\mu_2}{\sigma_2}{\alpha}
                }
                {e}
                {
                    \TypTrailMc{\tau_0}{\mu_0}{\sigma_0}{\tau'_0}{\mu'_0}{\sigma'_0}{\beta}
                }
          \end{array}
        }
        \\[1ex]
    \end{array}
$
}

\newcommand{\TypeCzVariable}{
$
    \begin{array}{c}
      \infer[\Rule{TControl0}]{ 
        \JudgeSTLC
            {\Gamma}
            {\ControlZ{k}{e}}
            {\TypTrailMc
              {\tau}
              {\mu_{\alpha}}{\sigma_{\alpha}}
              {\alpha} 
              {\mu_{\beta}}{(\Mc{\tau_0}{\mu_0}{\sigma_0}{\tau'_0}{\mu'_0}{\sigma'_0})}
              {\beta}} 
      }{ 
        \begin{array}{c}
          \Compatible
              {(\Arrow{\tau_1}{\mu_1}{\sigma_1}{\tau_2})}
              {\mu_2}
              {\mu_{\gamma}}
          \quad
          \Compatible
              {\mu_{\beta}}
              {\mu_{\gamma}}
              {\mu_{\alpha}}
          \\
          \JudgeSTLC
            {
              \Gamma,\,k\,:\,
              \TypTrailMc
                  {\tau\to \tau_1}
                  {\mu_1}{\sigma_1}
                  {\tau_2}
                  {\mu_2}{\sigma_{\alpha}}
                  {\alpha}
            }
            {e}
            {
              \TypTrailMc
                {\tau_0}
                {\mu_0}{\sigma_0}
                {\tau'_0}
                {\mu'_0}{\sigma'_0}
                {\beta}
            }
        \end{array}
      }
      \\[1ex]
    \end{array}
$
}

\newcommand{\TypePzVariable}{
$
    \infer[\Rule{TPrompt0}]{ 
      \JudgeSTLC
          {\Gamma}
          {\myReset{e}}
          {\TypTrailMc
            {\tau}
            {\mu_{\alpha}}{\sigma_{\alpha}}
            {\alpha}
            {\mu_{\beta}}{\sigma_{\beta}}
            {\beta}
          }
    }{ 
      \begin{array}{c}
        \IdContType
            {\gamma}{\mu_{id}}{\sigma_{id}}{\gamma'}
            \\
        \JudgeSTLC
            {\Gamma}
            {e}
            {\TypTrailMc
              {\gamma}
              {\mu_{id}}{\sigma_{id}}
              {\gamma'}
              {\bullet_{\mu}}{(\Mc{\tau}{\mu_{\alpha}}{\sigma_{\alpha}}{\alpha}{\mu_{\beta}}{\sigma_{\beta}})}
              {\beta}
            }
      \end{array}
    }
$
}

\newcommand{\TypePzVariablefDfun}{
$
    \infer[\Rule{4Dfun-Prompt0}]{ 
      \JudgeSTLCfDfun
          {\Gamma}
          {\myReset{e}}
          {\TypTrailMc
            {\tau}
            {\bullet_{\mu}}{\sigma_{\alpha}}
            {\alpha}
            {\bullet_{\mu}}{\sigma_{\alpha}}
            {\beta}
          }
    }{ 
      \begin{array}{c}
        \IdContType
            {\gamma}{\bullet_{\mu}}{\sigma_{id}}{\gamma'}
            \\
        \JudgeSTLCfDfun
            {\Gamma}
            {e}
            {\TypTrailMc
              {\gamma}
              {\bullet_{\mu}}{\sigma_{id}}
              {\gamma'}
              {\bullet_{\mu}}{(\tau\to\alpha)}
              {\beta}
            }
      \end{array}
    }
$
}

\newcommand{\TypePzVariablefDdash}{
$
    \infer[\Rule{TPrompt0$'$}]{ 
      \JudgeSTLC
          {\Gamma}
          {\myReset{e}}
          {\TypTrailMc
            {\tau}
            {\bullet_{\mu}}{\sigma_{\alpha}}
            {\alpha}
            {\bullet_{\mu}}{\sigma_{\alpha}}
            {\beta}
          }
    }{ 
      \begin{array}{c}
        \IdContType
            {\gamma}{\bullet_{\mu}}{\sigma_{id}}{\gamma'}
            \\
        \JudgeSTLC
            {\Gamma}
            {e}
            {\TypTrailMc
              {\gamma}
              {\bullet_{\mu}}{\sigma_{id}}
              {\gamma'}
              {\bullet_{\mu}}{(\Mc{\tau}{\bullet_{\mu}}{\sigma_{\alpha}}{\alpha}{\bullet_{\mu}}{\sigma_{\alpha}})}
              {\beta}
            }
      \end{array}
    }
$
}

\newcommand{\IdconttypeCompatible}{
$
  \begin{array}{lcl}
    \\[-1ex]
    \hspace{2cm}
    \IdContType{\tau}{\bullet_{\mu}}{\bullet_{\sigma}}{\tau'} & = & \tau \equiv \tau' \\
    \hspace{2cm}
    \IdContType{\tau}{\bullet_{\mu}}{(\Mc{\tau_1}{\mu_1}{\sigma_1}{\tau'_1}{\mu_2}{\sigma_2})}{\tau'} & = &
    (\tau \equiv \tau_1) \wedge (\tau' \equiv \tau'_1) \wedge (\mu_1 \equiv \mu_2) \wedge (\sigma_1 \equiv \sigma_2) \\
    \hspace{2cm}
    \IdContType{\tau}{(\Arrow{\tau_1}{\mu_1}{\sigma_1}{\tau'_1})}{\sigma_2}{\tau'} & = &
    (\tau \equiv \tau_1) \wedge (\tau' \equiv \tau'_1) \wedge (\mu_1 \equiv \bullet_{\mu}) \wedge (\sigma_1 \equiv \sigma_2)
    \\[1.2em]
    \Compatible{\bullet_{\mu}}{\mu_2}{\mu_3} & = & \mu_2 \equiv \mu_3 \\
    \Compatible{(\Arrow{\tau_1}{\mu_1}{\sigma_1}{\tau'_1})}{\bullet_{\mu}}{\mu_3} & = &
    (\Arrow{\tau_1}{\mu_1}{\sigma_1}{\tau'_1}) \equiv \mu_3 \\
    \Compatible
       {(\Arrow{\tau_1}{\mu_1}{\sigma_1}{\tau'_1})}
       {(\Arrow{\tau_2}{\mu_2}{\sigma_2}{\tau'_2})}{\bullet_{\mu}} & = & \bot \\
    \Compatible
        {(\Trail{\tau_1}{\mu_1}{\sigma_1}{\tau'_1})}
        {(\Trail{\tau_2}{\mu_2}{\sigma_2}{\tau'_2})}
        {(\Trail{\tau_3}{\mu_3}{\sigma_3}{\tau'_3})} & = &
    (\tau_1 \equiv \tau_3) \wedge (\tau'_1 \equiv \tau'_3) \wedge (\sigma_1 \equiv \sigma_3) \\[-0.1em]
    & &
    \quad \wedge (\Compatible{(\Arrow{\tau_2}{\mu_2}{\sigma_2}{\tau'_2})}{\mu_3}{\mu_1})\\
  \end{array}
$
}

\newcommand{\InferCzInterpVariable}{
  \[
    \begin{array}{c}
    \ECPS{\ControlZ{k}{e}} \,\rho\,
    \kappa^{\Rmath{(\Arrow{\tau}{\mu_{\alpha}}{\sigma_{\alpha}}{\alpha})^*}}  
    \,=\,\hfill\\
    \quad
    \Abs{t^{\Rmath{\mu_\beta^*}}}{}
    \Abs{((\kappa_0^{\Rmath{(\Arrow{\tau_0}{\mu_0}{\sigma_0}{\tau'_0})^*}},\,t_0^{\Rmath{{\mu'_0}^*}}),\,m_0^{\Rmath{{\sigma'_0}^*}})}{}\hfill \\
    \quad\quad
    (\ECPS{e}\,
    \rho[
      \Abs{v^{\Rmath{\tau^*}}}{}
      \Abs{\kappa'^{\Rmath{(\Arrow{\tau_1}{\mu_1}{\sigma_1}{\tau_2})^*}}}{}
      \Abs{t'^{\Rmath{\mu_2^*}}}{}
      \Abs{m'^{\Rmath{\sigma_2^*}}}{}\hfill\\
      \quad\quad\quad
      (\kappa^{\Rmath{(\Arrow{\tau}{\mu_{\alpha}}{\sigma_{\alpha}}{\alpha})^*}}
      \,v^{\Rmath{\tau^*}}\,\hfill\\
      \quad\quad\quad\quad
      (t^{\Rmath{\mu_\beta^*}}\,@\,
      (\kappa'^{\Rmath{(\Arrow{\tau_1}{\mu_1}{\sigma_1}{\tau_2})^*}} ::
      t'^{\Rmath{\mu_2^*}})^{\Rmath{\mu_\gamma^*}})^{\Rmath{\mu_\alpha^*}}\,
      m'^{\Rmath{\sigma_{\alpha}^*}})^{\Rmath{\alpha^*}}
      \,/\, k
    ] \\
    \hfill
    \kappa_0^{\Rmath{(\Arrow{\tau_0}{\mu_0}{\sigma_0}{\tau'_0})^*}} \,
    t_0^{\Rmath{{\mu'_0}^*}} \,
    m_0^{\Rmath{{\sigma'_0}^*}})^\Rmath{\beta^*}\\[1em]          
    \end{array}
  \]
}

%% file: abstract.tex
\begin{abstract}

The operational behavior of control operators has been studied comprehensively in the past few decades,
but type systems of control operators have not.  
There are distinct type systems for $\ttS$, $\ttC$, and $\ttSZ$ without
any relationship between them, and there has not been a type system 
that directly corresponds to $\ttCZ$.
This paper remedies this situation by giving a uniform type system for
all the four control operators.  Following Danvy and Filinski's
approach, we derive a monomorphic type system from the CPS interpreter that
defines the operational semantics of the four control operators.
By implementing the typed CPS interpreter in Agda, we show that the
CPS translation preserves types and that the calculus with all the
four control operators is terminating.  Furthermore, we show
the relationship between our type system and the previous type systems for
$\ttS$, $\ttC$, and $\ttSZ$.
\end{abstract}

%% file: intro.tex
\section{Introduction}

A continuation represents the remainder of computation.
The idea of continuations is versatile
since they can be used to suspend and resume computation.
There are many applications such as concurrent programming \cite{parallel-cml} 
and web programming \cite{krishnamurthi2007implementation}.

To make continuations accessible for programmers,
various control operators have been proposed to handle continuations
in direct-style programs, starting from the $\ttC$ and
$\ttP$ by
Felleisen \cite{Felleisen1988Prompt} and followed by
$\ttSR$ by Danvy and Filinski \cite{Danvy1990Control},
$\ttSRZ$ \cite{Danvy1989Functional}, and\\
$\ttCPZ$ \cite{Gunter1995Generalization}.
These operators can be largely classified into two categories, 
based on whether the range of computation that a control operator can capture
is dynamic or static.

As an alternative to control operators, Plotkin and Pretnar \cite{plotkin2009handlers}
have proposed algebraic effect handlers to control the flow of a program.
In recent years, this research area has been actively studied,
and we have witnessed many efforts such as building a type system 
for algebraic effect handlers \cite{bauer2013effect}.
Furthermore, it also turns out that there is a close relationship 
with delimited control operators.
For example, $\ttSZ$ \cite{Danvy1989Functional} is related to deep effect handlers \cite{Forster2017Expressive, pirog2019typed} and $\ttCZ$ \cite{Gunter1995Generalization} is related to shallow effect handlers.

In such a situation, there is an increasing need to clarify 
the type-level relationship between delimited control operators and 
algebraic effect handlers.
To this end, we first need a foundation that allows us to discuss 
delimited control operators at the type level,
which we present in this paper.

The contributions of this paper are as follows:
\begin{itemize}
    \item 
        We derive a monomorphic typing rule for $\ttCZ$ (that allows
        answer type modification) from the corresponding CPS interpreter.
    \item
        We scale this derivation method to derive a whole type system for $\ttS$, $\ttSZ$, $\ttC$, and $\ttCZ$. 
        Even though some of these operators have different dynamic behavior,
        we carefully design the type system to have a unified notation.
    \item 
        We formalize the underlying CPS interpreter and the type system using the proof assistant Agda \cite{NorellThesis}.
    \item
        We clarify the relationship between the type systems between the existing studies and ours. 
        In particular, we show that our type system subsumes the existing type systems.
        Some of the relationships are formalized in Agda.
\end{itemize}

The rest of the paper is organized as follows.
We begin by introducing the four delimited control operators
and the differences between them in \Sec{\ref{secDelimited}},
and show the syntax and semantics of the underlying language in \Sec{\ref{secDsCps}}.
Then, we present our type system for four delimited control operators in \Sec{\ref{secType}}, 
and compare the previous studies with ours in \Sec{\ref{secOtherTypeSystems}}.
Finally, we discuss related work in \Sec{\ref{secRelated}} and conclude in \Sec{\ref{secConclusion}}.
We provide the Agda code that is mentioned in the paper as
supplementary material\footnote{
        The implementation is accessible from \url{https://github.com/chiaki-i/type4d}.
}.

%% file: delimited-control.tex
\section{Delimited Control Operators}\label{secDelimited}
In this section, we introduce the four delimited control operators:
$\ttS$ ($\Shift{k}{e}$), $\ttC$ ($\Control{k}{e}$), $\ttSZ$ ($\ShiftZ{k}{e}$), and $\ttCZ$ ($\ControlZ{k}{e}$).
Each of them captures the continuation up to the nearest surrounding
delimiter $\texttt{reset}$ ($\myReset{e}$).
For example, $\ttS$ captures the continuation as shown below:
\begin{equation*}
    \myReset{(\Shift{k}{k\,(k\,2)})+3}+4 \, = \, \myReset{k\,(k\,2)\,[\Abs{x}{\myReset{x+3}}/k]}+4 \,=\,12
\end{equation*}
Once $\ttS$ captures its surrounding context up to the nearest delimiter, 
it is bound to the variable $k$, and we proceed to evaluate the body of $\ttS$.

The four operators are different in how they capture the current context 
and how they treat captured contexts.
Below shows the behavior of each operator, where $E[\dots]$ is an
evaluation context that does not enclose the hole with $\ttReset$.
\[
    \begin{array}{rcl}
        \myReset{E[\Shift{k}{e}]}    & = & \langle e\,[\Abs{x}{\langle E[x] \rangle}/k] \rangle \\
        \myReset{E[\Control{k}{e}]}  & = & \langle e\,[\Abs{x}{\Wmath{\langle} E[x] \Wmath{\rangle}}/k] \rangle \\
        \myReset{E[\ShiftZ{k}{e}]}   & = & \Wmath{\langle} e\,[\Abs{x}{\langle E[x] \rangle}/k] \Wmath{\rangle} \\ 
        \myReset{E[\ControlZ{k}{e}]} & = & \Wmath{\langle} e\,[\Abs{x}{\Wmath{\langle} E[x] \Wmath{\rangle}}/k] \Wmath{\rangle} \\
    \end{array}
\]
Unlike $\ttS$, the surrounding delimiter of $\ttSZ$ disappears
after it captures the continuation.
Then again, the continuation $k$ captured by $\ttC$ 
is not surrounded by a delimiter.
Both differences appear in the case of $\ttCZ$.

These differences become critical if the delimited control operators are nested, 
or the operators are included in the captured context.
In the following example, the range of the context that $\ttC$ captures is not
surrounded by $\ttReset$, unlike $\ttS$ and $\ttSZ$.
Since the captured context $k_1$ does not have any enclosing delimiters, 
$\Control{k_2}{\,\dots}$ in $k_1$ captures not only the original
context around it (i.e., $y+[\,]$) but also the invocation context
$2+[\,]$ of $k_1$.
\[
\begin{array}{rl}
      & \myReset{(\Control{k_1}{(2+k_1\,1)})+\Control{k_2}{(4+k_2\,3)}} \\
    = & \myReset{(2+k_1\,1) \, [\Abs{y}{y+\Control{k_2}{(4+k_2\,3)}}/k_1]}\\
    = & \myReset{2+(1+\Control{k_2}{(4+k_2\,3)})}\\
    = & \myReset{(4+k_2\,3)\,[\Abs{x}{2+(1+x)}/k_2]}\\
    = & \myReset{4+(2+(1+3))} = 10
\end{array}
\]
By executing more $\ttC$'s in captured continuations, $\ttC$ would
capture more invocation contexts.
We refer to this list of invocation contexts as a ``trail'' in this paper.
To define a typing rule for $\ttC$,
it is necessary to keep track of the type of trails.

Similarly, in the example below, the outer $\ttSZ$ operator does not have
access to the context outside the inner $\texttt{reset}$.
However, after the evaluation of the first $\ttSZ$,
the second $\ttSZ$ has access to the outer context
because the inner delimiter is removed by the first $\ttSZ$.
\[
\begin{array}{rl}
    & \myReset{\myReset{(\ShiftZ{k_1}{\ShiftZ{k_2}{e}})+2}+3}\\
    = & \myReset{(\ShiftZ{k_2}{e\,[\Abs{x}{\myReset{x+2}}/k_1]})+3}\\
    = & e\,[\Abs{x}{\myReset{x+2}}/k_1,\Abs{x}{\myReset{x+3}}/ k_2]
\end{array}
\]
By nesting more $\ttSZ$'s, we have access to more outer contexts.
We refer to the continuations of the surrounding context as
a ``meta continuation''.
To define a typing rule for $\ttSZ$,
it is necessary to keep track of the type of meta continuations.

%% file: ds-cps.tex
\begin{figure*}[t]
  \lambdaDSyntax
  \caption{Syntax of \ld}
  \label{figLambdaD}
\end{figure*}

\begin{figure}[t]
  \lambdaCSyntax
  \caption{Syntax of \lc}
  \label{figLambdaC}
\end{figure}

\section{Syntax and Semantics}\label{secDsCps}
In this section, we introduce languages \ld~and \lc.
The language \ld~ is a call-by-value, left-to-right,
$\lambda$-calculus extended with numbers and four delimited control operators.
The semantics of \ld~ is defined by a CPS interpreter 
which takes \ld~ as a source language, and translates it into \lc~,
which is a standard simply-typed $\lambda$-calculus extended with unit
and pairs to support meta continuations.
This CPS interpreter becomes the basis of 
our type system in \Sec{\ref{secType}}.

\subsection{Syntax of \ld}\label{subsecLambdaD}

\Fig{\ref{figLambdaD}} shows the syntax of \ld.
It is a call-by-value, left-to-right, $\lambda$-calculus
extended with numbers and delimited control operators.
There are four control operators that capture continuations:
$\ttS\ (\ShiftX)$, $\ttC\ (\ControlX)$, $\ttSZ\ (\ShiftZX)$,
and $\ttCZ\ (\ControlZX)$.
These operators are paired with a delimiter called
$\ttR$, $\ttP$, $\ttRZ$, and $\ttPZ$, respectively,
all of which behave the same \cite{Kiselyov2005Remove}.
In this paper, we use a single delimiter $\myReset{\,}$ to represent
all the four delimiters above.\footnote{
We call $\myReset{\,}$ as $\ttReset$ but use \Rule{TPrompt0}
for the name of the typing rule.
}

Types in \ld~ (also in \Fig{\ref{figLambdaD}}) have rather complicated forms.
The type of a continuation is the key to understanding the types in \ld.
In our type system, it is represented as $\Arrow{\tau_1}{\mu}{\sigma}{\tau_2}$, 
where it takes a value of type $\tau_1$, 
a trail of type $\mu$, and 
a meta continuation of type $\sigma$ and returns a value of type $\tau_2$.
Having the type of a continuation in mind, a function has the type
$\TypTrailMc{\tau_1\to\tau_2}{\mu_{\alpha}}{\sigma_{\alpha}}{\alpha}{\mu_{\beta}}{\sigma_{\beta}}{\beta}$,
which corresponds to a standard function type $\tau_1\to\tau_2$ but is
evaluated with
\begin{itemize}
    \item a continuation of type $\Arrow{\tau_2}{\mu_{\alpha}}{\sigma_{\alpha}}{\alpha}$,
    \item a trail of type $\mu_{\beta}$, and 
    \item a meta continuation of type $\sigma_{\beta}$,
\end{itemize}
and finally results in a value of type $\beta$.

A trail and a meta continuation are internal representations of
delimited control operators, 
which explicitly appear only in the CPS interpreter shown in \Sec{\ref{subsecLambdaC}}.
However, their types are essential to define 
the type of the operators in \Sec{\ref{secType}}.
A type of a trail is either $\bullet_{\mu}$ meaning the trail is
empty or a continuation representing a composition of invocation contexts.
A type of a meta continuation is either $\bullet_{\sigma}$ meaning the
meta continuation is empty, or
a list of pairs of a continuation and a trail of surrounding contexts.

An ``answer type'' is the type of the surrounding context or the
return type of a continuation.
In \Fig{\ref{figLambdaD}}, an arrow type contains two answer types:
$\alpha$ is an initial answer type
and $\beta$ is a final answer type.
Initial answer type refers to the term's expected return type, while
final answer type is the term's actual return type.
Take the expression $\myReset{(\Abs{x}{\texttt{isZero}\,({x-1})})\,1}$ as an example.
Assuming that \texttt{isZero} is a function from \texttt{int} to
\texttt{bool},
both the initial and final answer types of the function $(\Abs{x}{\,\dots})$ is 
\texttt{bool}, because the context returns a boolean value.
On the other hand, in an expression 
$\myReset{(\Abs{x}{\texttt{isZero}\,(\Shift{k}{x-1})})\,1}$,
the initial answer type of the function (\texttt{bool}) changes to the
final one (\texttt{int}): although the value originally returned by
the context is a boolean (the return value of \texttt{isZero}),
the actual return values is a number because of the $\ttS$.
Such a change in type is called Answer Type Modification (ATM),
and this can happen when we use delimited control operators.


\subsection{Syntax of \lc}\label{subsecLambdaC}

\Fig{\ref{figLambdaC}} shows the syntax of \lc,
the target language of CPS interpreter to be presented
in \Sec{\ref{subsecInterpreter}}.
Basically, it is a standard simply-typed $\lambda$-calculus extended
with a unit, a pair, and a case-analysis construct. 
We elaborate the types of \lc~at the end of \Sec{\ref{subsecInterpreter}}.

We use $()$ of type \texttt{unit} to represent an empty trail and a
continuation (of the form $\Abs{v}{\Abs{t}{\Abs{m}{\dots}}}$) to
represent a non-empty trail.
The same representation is used by Shan \cite{Shan2007Delimited}.
Representing a non-empty trail as a function enables us to have
invocation contexts of different types \cite{Cong2021Functional}.

We also use $()$ to represent an empty meta continuation.
A non-empty meta continuation is represented as a (heterogeneous) list
of pairs of a continuation and a trail, represented using $()$ as an
empty list and a pair as a cons.



\begin{figure*}[t]
    \CpsInterpreterColor
    \caption{CPS Interpreter for \ld}
    \label{figCPSInterpreter}
\end{figure*}

\begin{figure*}[t]
    \CpsTranslation
    \caption{Type-level CPS Translation for \ld}
    \label{figCPSType}
\end{figure*}

\subsection{CPS Interpreter}\label{subsecInterpreter}

We now define a CPS interpreter of \ld,
represented as $\ECPS{\,}$.
We begin by the definition of CPS interpreter shown in \Fig{\ref{figCPSInterpreter}}.
This interpreter is based on a 2CPS\footnote{
When a CPS expression is translated once more into CPS,
it is called 2CPS \cite{Danvy1990Control}. While a CPS expression carries a continuation,
a 2CPS expression carries both a continuation and a meta continuation.
To distinguish between these two styles, sometimes the standard CPS 
is called 1CPS.
}
interpreter \cite{Danvy1989Functional} receiving a continuation
$\kappa$ and a meta continuation $m$.
On top of them, it receives a trail $t$ to account for invocation
contexts.
This is the same as the 
interpreter shown by Shan \cite{Shan2007Delimited}.
In general, the interpreter $\ECPS{e}$ takes an environment $\rho$, 
a continuation $\kappa$, a trail $t$, and a meta continuation $m$ and
evaluates the term $e$.
The meta continuation $m$ and the trail $t$ do not play any roles for
the first four cases; if we $\eta$-reduce them, the interpreter
becomes a standard CPS interpreter.
In the interpreter, we highlight (important parts of) trails with underlines,
and meta continuations with wavy lines.
In the cases of delimited control operators,
it binds the captured continuation to a variable $k$, 
adds $k$ to the environment $\rho$,
and recursively evaluates $e$ with other arguments.

For $\ttS$ and $\ttC$, the body $e$ is evaluated with the initial continuation $\idk$,
an empty trail $()$, and a meta continuation $m$.
That is, the current continuation and trail are cleared and the meta
continuation remains unchanged, which means that the body $e$ is
evaluated in an empty context and is delimited by reset.
For example, 
$\myReset{(\Control{k_1}{e_1})+1}$ becomes $\myReset{e_1}$ 
with $k_1=\Abs{x}{x+1}$,
where the surrounding context $\Abs{x}{x+1}$ is cleared and the
reset around the original expression remains.

On the other hand, in the cases for $\ttSZ$ and $\ttCZ$, 
$e$ is evaluated with the continuation $\kappa_0$ and trail $t_0$ that
are found in the meta continuation.
That is, the current continuation and trail are taken from the ones
outside the current delimiter, which means that the body $e$ has
access to that information.
For example,
$\myReset{E[\myReset{(\ControlZ{k_1}{e_1})+1}]}$ becomes
$\myReset{E[e_1]}$
with $k_1=\Abs{x}{x+1}$,
where not only the surrounding context $\Abs{x}{x+1}$ is cleared but
also the delimiter is removed resulting $e_1$ to have access to the
context outside the original inner delimiter.
 

Next, let us focus on the structure of the captured continuation $k$ 
in each delimited control operator.
For $\ttS$ and $\ttSZ$, the then-current continuation and trail $(\kappa',\,t')$ 
are added to then-current meta continuation $m'$.
Since control operators capture the current continuation and trail,
not the ones in a meta continuation, it means that the captured
continuation $\kappa$ does not have direct access to those
continuation and trail (unless $\ttSZ$ or $\ttCZ$ is used to
deconstruct a meta continuation).
That is, the body of the captured continuation retains the surrounding
reset.

We use pairs to represent meta continuations 
to record every snapshot of $\kappa'$ and $t'$
because $\ttSZ$ and $\ttCZ$ need access to each layer of a meta continuation.
If we had only $\ttS$ and $\ttC$ as control operators, 
a meta continuation could be represented as a function
as seen in Danvy and Filinski's CPS interpreter for $\ttSR$ \cite{Danvy1989Functional}. 

For the captured continuation of $\ttC$ and $\ttCZ$,
on the other hand,
$\kappa'$ and $t'$ are appended to the trail $t$.
It means that the captured continuation $\kappa$ does have direct
access to the continuation and trail with $\ttS$ or $\ttC$
(without using $\ttSZ$ or $\ttCZ$ to deconstruct a meta continuation).
That is, the body of the captured continuation is not surrounded by
reset.

The $\_@\_$ operator appends two trails, 
while $\_\!::\!\_$ operator conses a continuation to a trail.
We use a function form to express both trails and continuations as mentioned in \Sec{\ref{subsecLambdaD}}.
Thus, both $\_@\_$ and $\_\!::\!\_$ 
are implemented as function composition\footnote{
The implementation of trail compositions
corresponds to the \texttt{compose} function 
in Shan's work \cite{Shan2007Delimited}.}
(\Fig{\ref{figCPSInterpreter}}).

Now that we introduced the CPS interpreter and its auxiliary functions, 
we take a close look at the type of these functions.
To type the case-analysis construct in the three auxiliary functions 
$\idk$, $\_@\_$, and $\_\!::\!\_$, 
it is necessary to check the type of the given arguments.
For example, $\idk$ takes $(t,\,m)$ and splits cases 
based on whether each of $t$ and $m$ has the type of $\texttt{unit}$.
In accordance with these functions, the CPS interpreter also
implicitly carries additional constraints on the types of $t$ and $m$\footnote{These constraints are later implemented as \textsf{id-cont-type} and \textsf{compatible} in \Fig{\ref{figLDTypeSystem}}. 
Also, the corresponding implementation in Agda precisely follows these definitions.}.

Finally, \Fig{\ref{figCPSType}} shows the CPS translation of types.
The type of a trail $\Arrow{\tau_1}{\mu}{\sigma}{\tau_2}$ in \ld~
is translated into a standard function type
with all arguments recursively translated.
The type of a meta continuation is a product,
which means a meta continuation is represented as a nested pair.


%% file: type-system.tex
\section{Type System of \ld}\label{secType}

\begin{figure*}[t]
    \centering
  \text{\fbox{$\JudgeSTLC{\Gamma}{e}{\TypTrailMc{\tau}{\mu_{\alpha}}{\sigma_{\alpha}}{\alpha}{\mu_{\beta}}{\sigma_{\beta}}{\beta}}$}} \hfill
  \vspace{-2em}
  \TypeLambdaVariable
  \TypeSVariable
  \TypeCVariable
  \TypeSzVariable
  \TypeCzVariable
  \TypePzVariable
  \IdconttypeCompatible
  \caption{Type System of $\lambda_{D}$}
  \label{figLDTypeSystem}
\end{figure*}

In this section, we introduce the type system of \ld~(\Fig{\ref{figLDTypeSystem}}).
Following Danvy and Filinski \cite{Danvy1989Functional} and
Cong et al.\ \cite{Cong2021Functional},
we derive this type system directly from the CPS interpreter in \Fig{\ref{figCPSInterpreter}}.

The typing judgment 
$\JudgeSTLC{\Gamma}{e}{\TypTrailMc{\tau}{\mu_{\alpha}}{\sigma_{\alpha}}{\alpha}{\mu_{\beta}}{\sigma_{\beta}}{\beta}}$ reads:
``under a type environment $\Gamma$, a term $e$ has the type $\tau$. 
When $e$ is evaluated with
a continuation of type $\Arrow{\tau}{\mu_{\alpha}}{\sigma_{\alpha}}{\alpha}$,
a trail of type $\mu_{\beta}$, and 
a meta continuation of type $\sigma_{\beta}$,
it eventually reduces to be a value of type $\beta$.''

Among all the typing rules in \Fig{\ref{figLDTypeSystem}}, the rules
\Rule{TShift0}, \Rule{TControl0}, and \Rule{TPrompt0} are completely new.
The rules \Rule{TShift} and \Rule{TControl} are 
based on the previous work \cite{Danvy1989Functional, Cong2021Functional}, 
but they are new in that they consider trails and meta continuations together.

\subsection{How to Derive the Typing Rules}\label{subsecTControl0}

In this section, we take \Rule{TControl0} as an example to show
how to derive the typing rule from the CPS interpreter. 
Below is the evaluation rule of $\ttCZ$ from \Fig{\ref{figCPSInterpreter}}
with type annotations on the upper right corner of each term.
\InferCzInterpVariable
All the annotated types follow the typing rules of the target language
\lc, once we expand the CPS translation of types $^*$.
From these types, we derive the typing rule for $\ControlZ{k}{e}$.

When we evaluate $\ControlZ{k}{e}$, it takes these parameters:
\begin{itemize}
    \item a continuation $\kappa$ of type $(\Arrow{\tau}{\mu_{\alpha}}{\sigma_{\alpha}}{\alpha})^*$, 
    \item a trail $t$ of type $\mu_{\beta}^*$, and 
    \item a meta continuation $((\kappa_0,\,t_0),\,m_0)$ of type\\
$(\Mc{\tau_0}{\mu_0}{\sigma_0}{\tau'_0}{\mu'_0}{\sigma'_0})^*$.
\end{itemize}
The rest of the term $(\ECPS{e}\,\dots)$ evaluates to a value of type $\beta^*$.
These types match the conclusion part of \Rule{TControl0}.
Notice that the types in the evaluation rule are all translated into CPS
following the rules in \Fig{\ref{figCPSType}}.
The same goes for the type of term $e$ and $k$.
We maintain exact correspondence between the type-annotated CPS
interpreter and the typing rules.

The only remaining part is the typing constraint of the trail composition
$(t\,@\,\kappa'::t')$, which we cover in the following section.

\subsection{Typing Constraints}\label{subsecConstraints}

In \Fig{\ref{figLDTypeSystem}}, there are two typing constraints called
\textsf{compatible} and \textsf{id-cont-type}.
The former is for trail compositions, and the latter is for the initial continuation.
Similarly to \Sec{\ref{subsecTControl0}}, these constraints are
directly derived from the corresponding definitions in \Fig{\ref{figCPSInterpreter}}.

The rules of \textsf{compatible} correspond to the trail composition
$\_@\_$ and $\_\!::\!\_$.
As mentioned in \Sec{\ref{subsecInterpreter}}, they are both implemented 
as function compositions.
When the trail types satisfies $\Compatible{\mu_1}{\mu_2}{\mu_3}$, 
it means that we have a trail $t_1$ of type $\mu_1$ and
a trail $t_2$ of type $\mu_2$, 
and they are composed to become a trail $t_3$ of type $\mu_3$.
Specifically, the first rule of \textsf{compatible} corresponds to 
$()\,@\,t\,=\,t$, the second one to $k::()\,=\,k$.
The third one says that the result of composing non-empty trails 
should not be empty.
The last one seems a little complicated, but this is derived from
the last rule:
\InferCompatible
At the final line of the equation above, $\tau_1^*$ should be equal to $\tau_3^*$
so that $k$ can receive the value $v$ of type $\tau_3^*$ as an argument.
Also, in order for $k$ to receive $(t::t')$ as the second argument, 
it is necessary to use \textsf{compatible} once more to say that
composing two trails of type $(\Arrow{\tau_2}{\mu_2}{\sigma_2}{\tau'_2})^*$ and 
$\mu_3^*$ should become a trail of type $\mu_1^*$.
These constraints correspond to the two \textsf{compatible}
constraints in the rule \Rule{TControl0}.

Moving on to the typing constraints of the initial continuation.
In \Fig{\ref{figCPSInterpreter}}, $\ttS$, $\ttC$, and $\ttPZ$ use
the initial continuation. 
Accordingly, the initial continuation of type $\Arrow{\gamma}{\mu_{id}}{\sigma_{id}}{\gamma'}$
needs to satisfy the constraint
$\IdContType{\gamma}{\mu_{id}}{\sigma_{id}}{\gamma'}$
in \Fig{\ref{figLDTypeSystem}}.
For example, the following shows how the second case of $\idk$ 
(in \Fig{\ref{figCPSInterpreter}})
is associated with types:
\InferIdContType
The $k_0$'s arrow type matches the types of its two arguments,
and the return type of $k_0$ matches the return type of $\idk$.

\subsection{Properties and Observations}

Both the CPS interpreter and the type system of \ld~ are formalized in
Agda in an intrinsically-typed way \cite{altenkirch-monadic}
using PHOAS \cite{Chlipala2008}.
Namely, all the terms are typed by construction and thus, successfully
implementing the CPS interpreter as a well-typed Agda program means
that CPS translation preseves types.

\begin{theorem}[Type Preservation of CPS Translation]
\label{thmTypePreservation}
    ~\\
    If $~\JudgeSTLC{\Gamma}{e}{\TypTrailMc{\tau}{\mu_{\alpha}}{\sigma_{\alpha}}{\alpha}{\mu_{\beta}}{\sigma_{\beta}}{\beta}}$ in \ld, then
    for any $\rho$ that respects $\Gamma^*$ (i.e.,
    $\JudgeSTLC{}{\rho(x)}{\Gamma^*(x)}$ for any $x$),
    $\JudgeSTLC{\Gamma^*}{\ECPS{e} \,\rho}{(\tau^* \to \mu_{\alpha}^* \to \sigma_{\alpha}^* \to \alpha^*) \to \mu_{\beta}^* \to \sigma_{\beta}^* \to \beta^*}$ in \lc.
\end{theorem}

Furthermore, since the CPS interpreter passes Agda termination
checker, we have that the evaluation of the \ld~ term always
terminates, given the initial continuation, trail, and meta
continuation.

\begin{theorem}[Termination]
\label{thmTermination}
    ~\\
    If $~\JudgeSTLC{}{e}{\TypTrailMc{\tau}{\bullet_{\mu}}{\bullet_{\sigma}}{\tau}{\bullet_{\mu}}{\bullet_{\sigma}}{\tau}}$ in \ld,
    then evaluation of $e$ terminates.\\
\end{theorem}

%% file: other-systems.tex
\section{Comparisons with Other Type Systems}\label{secOtherTypeSystems}

So far, we explained our type system for four delimited control operators
(referred to below as \sffD).
In this section, we compare ours with other type systems presented in previous research.

\subsection{$\ttSR$}\label{subsecSR}

In \Sec{\ref{secType}}, we derived our type system from a CPS interpreter
following Danvy and Filinski's approach \cite{Danvy1989Functional}.
This section compares their type system for $\ttSR$ (referred to below as \sfDF) 
with \sffD.

Here is the overview diagram of the relationship between the underlying interpreters of $\sfDF$ and $\sffD$.
The solid lines in the diagram show that the corresponding translation
is formalized in Agda.
\begin{equation*}
\begin{tikzcd}[column sep=large]
\textrm{2CPS}
  &[-2ex]
\sfDFtwo 
    \arrow[r,"\textit{Add trail}", shift left=0.6ex]
    \arrow[d,"\textit{Direct Style}", shift left=0.6ex]
  &[2ex]
\sffDfun 
    \arrow[l,"\textit{Remove trail}", shift left=0.6ex]
    \arrow[r,dotted,"\textit{Defunctionalize}",shift left=0.6ex]
  &[5ex]
\sffD \arrow[l,dotted,"\textit{Functionalize}", shift left=0.6ex]
  \\
\textrm{1CPS}
  &[-2ex]
\sfDF \arrow[u,"\textit{CPS}", shift left=0.6ex]
  &[2ex]{} &[5ex] {}
\end{tikzcd}
\end{equation*}
There are three differences between $\sfDF$ and $\sffD$: 
whether the underlying CPS interpreter is written in 1CPS or 2CPS,
whether the interpreter takes a trail as an additional parameter, and
whether the meta continuation's data structure is a function or
a nested pair.
If we translate the CPS interpreter for $\sfDF$ once more into 2CPS,
we get $\sfDFtwo$'s interpreter.
After that, if we add trails to $\sfDFtwo$'s,
it becomes $\sffDfun$'s.
Finally, if we defunctionalize the meta continuations in $\sffDfun$'s, 
we get $\sffD$'s.

In the following discussion, we take a closer look at 
the relationship between type systems:
A. $\sfDF$ and $\sfDFtwo$, B. $\sfDFtwo$ and $\sffDfun$,
and C. $\sffDfun$ and $\sffD$.
The goal of this section is to show that 
a term $e$ is typable in $\sfDF$ if and only if $e$ is typable in $\sffD$.
Note that the source language for all these interpreters only contains
simply-typed lambda calculus and $\ttSR$.

\begin{figure*}[t]
    \TypeSRDFTrans 
    \caption{Type-level translation between $\sfDF$ and $\sfDFtwo$}
    \label{figSRTransA1}
\end{figure*}

\begin{figure*}
    \TypeSRTrailTrans 
    \caption{Type-level translation between $\sfDFtwo$ and $\sffDfun$}
    \label{figSRTransB1}
\end{figure*}

\paragraph{A. $\sfDF$ and $\sfDFtwo$}
The underlying interpreter of \sfDF~ is 1CPS, 
which means it does not contain any meta continuations.
Also, this interpreter does not contain trails 
since it is not necessary unless the source language contains $\ttC$ or $\ttCZ$.
The $\sfDF$ typing judgment $\Gamma\vdash_{\textsf{DF}}\,e\,:\,\tau,\,\alpha,\,\beta$
reads that a $\sfDF$ term $e$ has type $\tau$, 
and evaluating $e$ changes the answer type from $\alpha$ to $\beta$. 
Below is the CPS interpreter and the typing rule \cite{Danvy1989Functional}
for $\ttS$ as an example.
\CpsInterpreterDF
\TypeSDFWoTrailMc
The type of functions (as witnessed by the type of $k$)
has the form $\tau_1\to\tau_2,\,\alpha,\,\beta$
which is a function from $\tau_1$ to $\tau_2$ and the body of the
function changes the answer type from $\alpha$ to $\beta$.
The interpreter and the typing judgment are simpler than those of $\sffD$
because unlike other control operators, 
$\ttS$ does not change invocation contexts or meta contexts.

On the other hand, the typing judgment of $\sfDFtwo$ looks similarly to $\sffD$;
$\Gamma\vdash_{\textsf{DF-2}}\,e\,:\,\TrailMC{\tau}{\sigma_{\alpha}}{\alpha}{\sigma_
{\beta}}{\beta}$ means $e$ has type $\tau$, and it results in a value of type $\beta$ 
if $e$ is evaluated with a continuation of type $\tau\to\langle\sigma_{\alpha}\rangle\,\alpha$ 
and the meta continuation of type $\sigma_{\beta}$.
Below is the CPS interpreter (which is identical to the one presented
by Danvy and Filinski \cite{Danvy1990Control})
and the typing rule for $\ttS$ derived from the $\sfDFtwo$ interpreter.
In $\sfDFtwo$, a meta continuation has a function type.
\CpsInterpreterDFFunc
\TypeSDFFunc
Now that we have these two type systems, $\sfDF$ and $\sfDFtwo$,
we show that a term is typable in $\sfDF$ if and only if it is typable
in $\sfDFtwo$.
\Fig{\ref{figSRTransA1}} describes the type-level translations.
The first translation ($\DFonetwo{\tau}{\gamma}$) is for a $\sfDF$ type,
and it translates a $\sfDF$ type $\tau$ recursively
into a $\sfDFtwo$ type with an arbitrary but fixed $\sfDFtwo$ answer type $\gamma$.
The second one ($\DFonetwogamma{\Gamma}$) is for a $\sfDF$ type environment,
where $\bullet_{\Gamma}$ represents an empty type environment.
Similarly, we define the reverse translation ($\DFtwoone{\tau}$ and $\DFtwoonegamma{\Gamma}$)
that strips off the answer type.
The type-level translations intuitively mean that
the answer type of meta continuations does not change,
because $\ttS$ does not have access to meta continuations.
With these translations, we can show that typability of $\sfDF$ and
$\sfDFtwo$ is the same.
\begin{theorem}[Typability between $\sfDF$ and $\sfDFtwo$]
\label{thmSROneTwo}
    ~\\
    (1)
    ~If $~\Gamma\,\vdash_{\textsf{DF}}\,e\,:\,\tau,\,\alpha,\,\beta$, 
    then for any $\sfDFtwo$ type $\gamma$, \\
    $\DFonetwogamma{\Gamma}\vdash_{\textsf{DF2}}\,e\,:\,\TrailMC{\DFonetwo{\tau}{\gamma}}{\DFonetwo{\alpha}{\gamma}\to\gamma}{\gamma}{\DFonetwo{\beta}{\gamma}\to\gamma}{\gamma}$.\\
    (2)
    ~If $~\Gamma\,\vdash_{\textsf{DF}}\,e\,:\TrailMC{\tau}{\alpha\to\gamma_1}{\gamma_2}{\beta\to\gamma_3}{\gamma_4}$, 
    then,\\
    $\DFtwoonegamma{\Gamma}\vdash_{\textsf{DF2}}\,e\,:\,\DFtwoone{\tau},\,\DFtwoone{\alpha},\,\DFtwoone{\beta}$.
    $\hfill\qed$
\end{theorem}
The proof is by induction on the typing derivation and is
formalized in Agda.

\begin{figure*}
    \TypeSRTransTofun
    \caption{Type-level translation from $\sffD$ to $\sffDfun$
    (that does not work)}
    \label{figSRTransC2}
\end{figure*}

\paragraph{B. $\sfDFtwo$ and $\sffDfun$}
Next, we move on to the relationship between $\sfDFtwo$ and $\sffDfun$.
The $\sffDfun$'s interpreter and its initial continuation $\idk''$ take a trail
as an additional argument compared to $\sfDFtwo$.
Plus, the typing judgment has the form
$\Gamma\,\vdash_{\textsf{4D-fun}}\,e\,:\,\TypTrailMc{\tau}{\mu_{\alpha}}{\sigma_{\alpha}}{\alpha}{\mu_{\beta}}{\sigma_{\beta}}{\beta}$, 
which reads exactly the same as $\sffD$.
Note that all the trails in \Rule{4Dfun-Shift} are empty,
because $\ttS$ does not modify any trails during evaluation.
\CpsInterpreterFDFunc
\TypeSfDWoTrailMc

The type-level translations from $\sfDFtwo$ to $\sffDfun$
($\DFtwofun{\tau}$, $\DFtwofunmeta{\sigma}$, and $\DFtwofungamma{\Gamma}$) 
and from $\sffDfun$ to $\sfDFtwo$ 
($\DFfuntwo{\tau}$, $\DFfuntwometa{\sigma}$, and $\DFfuntwogamma{\Gamma}$) 
are defined in \Fig{\ref{figSRTransB1}}.
We simply add empty trails or remove trails of the function type
because trails are not necessary for $\ttS$
and remain unused in $\sffDfun$.
With these type-level translations, we can show the typability of
$\sfDFtwo$ and $\sffDfun$ is the same.

\begin{theorem}[Typability between $\sfDFtwo$ and $\sffDfun$]
\label{thmSRTwoFun}
~\\
    (1) If $~\Gamma\,\vdash_{\textsf{DF2}}\,e\,:\,\TrailMC{\tau}{\sigma_{\alpha}}{\alpha}{\sigma_{\beta}}{\beta}$, then \\
    $\DFtwofungamma{\Gamma}\,\vdash_{\textsf{4Dfun}}\,e\,:\,\TypTrailMc{\DFtwofun{\tau}}{\bullet_{\mu}}{\DFtwofunmeta{\sigma_{\alpha}}}{\DFtwofun{\alpha}}{\bullet_{\mu}}{\DFtwofunmeta{\sigma_{\beta}}}{\DFtwofun{\beta}}$.\\
    (2) If
    $~\Gamma\,\vdash_{\textsf{4Dfun}}\,e\,:\,\TypTrailMc{\tau_1\to\tau_2}{\mu_1}{\sigma_1}{\alpha}{\mu_2}{\sigma_2}{\beta}$, then\\ $\DFfuntwogamma{\Gamma}\,\vdash_{\textsf{DF2}}\,e\,:\,\TrailMC{\DFfuntwo{\tau_1}\to\DFfuntwo{\tau_2}}{\DFfuntwometa{\sigma_1}}{\DFfuntwo{\alpha}}{\DFfuntwometa{\sigma_2}}{\DFfuntwo{\beta}}$.
    \hfill$\qed$
\end{theorem}
The proof is again by induction on the typing derivation and is
formalized in Agda.

\begin{figure*}
    \TypeCPTransOneTwo
    \caption{Type-level translation from $\sfCP$ to $\sffDfun$}
    \label{figTransCPonetwo}
\end{figure*}

\paragraph{C. $\sffDfun$ and $\sffD$}
Lastly, we investigate the relationship between $\sffDfun$ and $\sffD$.
The underlying interpreter of $\sffD$ is a variant of $\sffDfun$
where $\sffDfun$'s meta continuations are defunctionalized.

Below is the $\sffD$'s evaluation rule (taken from
\Fig{\ref{figCPSInterpreter}}) and a slightly modified typing rule for
$\ttS$.
\CPSInterpreterFDShift
\TypeSDashVariable
In \Rule{TShift$'$}, notice that the types in the underlined part 
are the same as those in the wavy-lined part (unlike the original
\Rule{TShift} in \Fig{\ref{figLDTypeSystem}}).
This is because the source language for $\sffD$ in this section
contains only $\ttS$ as a delimited control operator.
Evaluating $\ttS$ does not change meta continuations
unlike $\ttSZ$ or $\ttCZ$.

The interpreters for $\sffDfun$ and $\sffD$ receive the same trails
and meta continuations,
and so do the captured continuation $k$ in both interpreters.
Correspondingly, each of typing rules in $\sffDfun$ and $\sffD$ 
has the same type parameters.
The only difference is whether the trail type and the meta-continuation type are 
explicitly stored in a meta continuation.
Concretely, \Rule{4Dfun-Shift} contains type $\mu_1$ and $\sigma_1$ in $k$,
but not explicitly in the meta-continuation type $(\tau_1\to\tau_2)$ in $\Shift{k}{e}$.
On the other hand, the meta-continuation type of $\Shift{k}{e}$
in \Rule{TShift$'$} contains type $\bullet_{\mu}$ and $\sigma_1$, shown as 
$((\tau_1\to\langle\bullet_{\mu},\,\sigma_1\rangle\,\tau_2)\,\times\,\bullet_{\mu}::\sigma_1)$.

Now it is clear that we can convert each typing rule in $\sffDfun$ to
the one in $\sffD$ (and vice versa) by adding (deleting) the trail and
meta-continuation types in the premise to (from) the meta-continuation
type at the conclusion.
The same holds for all the other typing rules (see the Appendix).
Thus, a term $e$ is typable in $\sffD$ if and only if it is typable in
$\sffDfun$, proving the following theorem.
\begin{theorem}[Typability between $\sffDfun$ and $\sffD$]
\label{thmfDtofun1}
    ~\\
    (1) If $~\Gamma\,\vdash_{\textsf{4Dfun}}\,e\,:\,\TypTrailMc{\tau}{\mu_1}{\sigma_1}{\alpha}{\mu_2}{\sigma_2}{\beta}$, then\\
    $~\Gamma\,\vdash_{\textsf{4D}}\,e\,:\,\TypTrailMc{\tau'}{\mu'_1}{\sigma'_1}{\alpha'}{\mu'_2}{\sigma'_2}{\beta'}$ for some
    $\tau'$, $\mu'_1$, $\sigma'_1$, $\alpha'$, $\mu'_2$,
    $\sigma'_2$, and $\beta'$.\\
    (2) If $~\Gamma\,\vdash_{\textsf{4D}}\,e\,:\,\TypTrailMc{\tau}{\mu_1}{\sigma_1}{\alpha}{\mu_2}{\sigma_2}{\beta}$, then\\
    $~\Gamma\,\vdash_{\textsf{4Dfun}}\,e\,:\,\tau'
    \langle\mu_1',\,\sigma_1'\rangle\,$
    $\alpha' \langle\mu_2',\,\sigma_2'\rangle\,\beta'$ for some
    $\tau'$, $\mu'_1$, $\sigma'_1$, $\alpha'$, $\mu'_2$,
    $\sigma'_2$, and $\beta'$.\\
    \hfill$\qed$
\end{theorem}

We have not formalized this theorem in Agda, because for the first
part, it is not clear
how to define the type-level translation.
A simple translation that adds arbitrary $\mu$ and $\sigma$ to
$\sffDfun$'s non-empty meta continuation $(\tau_1\to\tau_2)$ to 
create $\sffD$'s non-empty meta continuation
$(\Mc{\tau_1}{\mu}{\sigma}{\tau_2}{\mu}{\sigma})$
does not work, because it causes conflicts in \Rule{TShift$'$}.
In the \Rule{TShift$'$} case,
it is necessary to show that $\Shift{k}{e}$ has type
$\tau\,\langle\mu_{\beta},\,\Mc{\tau_1}{\mu}{\sigma}{\tau_2}{\mu}{\sigma}\rangle\,\alpha\,\langle\mu_{\beta},\,\sigma_{\beta}\rangle\,\beta$,
but we only have variable $k$ of type 
$(\TypTrailMc{\tau\to\tau_1}{\bullet_{\mu}}{\sigma_1}{\tau_2}{\bullet_{\mu}}{\sigma_1}{\alpha})$,
where types $\bullet_{\mu}$ and $\sigma_1$ in $k$ do not match
the types $\mu$ and $\sigma$ in the goal of this case.

For the second part of the theorem, it is tempting to define a
type-level translation as in \Fig{\ref{figSRTransC2}}, where the empty
meta continuation $\bullet_{\sigma}$ is translated to
$\gamma \to \gamma$ for a given $\gamma$.
This translation does not work, however, because uniformly translating
$\bullet_{\sigma}$ to the same $\gamma \to \gamma$ turns out to
conflict with the constraint \textsf{id-cont-type}.
To define precise correspondence, we need to fine tune the translation
of $\bullet_\sigma$ that amounts to comparing the type derivations in
$\sffD$ and $\sffDfun$.

\subsection{$\ttCP$}\label{subsecCP}

Cong et al.~\cite{Cong2021Functional}
derived a type system for \texttt{control} \texttt{/prompt}
(referred to below as $\sfCP$)
from a corresponding 1CPS interpreter. 
The relationship between $\sfCP$ and $\sffD$ is as follows.
Since $\sfCP$ already contains trails to handle invocation contexts,
the diagram below is more straightforward than that of \Sec{\ref{subsecSR}}.
The difference between $\sfCP$ and $\sffD$ is
whether the underlying interpreter is 1CPS or 2CPS,
and whether the meta continuations are represented as functions or pairs.
\begin{equation*}
\begin{tikzcd}[column sep=large]
\textrm{2CPS}
  &[-2ex]
\sffDfun 
    \arrow[r,dotted,"\textit{Defunctionalize}",shift left=0.6ex]
    \arrow[d,dotted,"\textit{Direct Style}", shift left=0.6ex]
  &[5ex]
\sffD \arrow[l,dotted,"\textit{Functionalize}", shift left=0.6ex]
  \\
\textrm{1CPS}
  &[-2ex]
\sfCP \arrow[u,"\textit{CPS}",shift left=0.6ex]
  &[5ex] {}
\end{tikzcd}
\end{equation*}

The translation from $\sfCP$ to $\sffDfun$ is defined
as $\CPonetwosuper{*}$
in \Fig{\ref{figTransCPonetwo}}.
The translation rules are a variant of the first half of
\Fig{\ref{figSRTransA1}}.
Based on these rules, we show that a $\sfCP$ term is typable in $\sffDfun$,
and vice versa.

\begin{theorem}[Typability between $\sfCP$ and $\sffDfun$]
\label{thmCPfDfun}
    ~\\
    (1) If 
    $~\Gamma\,\vdash_{\textsf{CP}}\,e\,:\,\TrailMC{\tau}{\mu_1}{\alpha}{\mu_2}{\beta}$, 
    then for any $\sffDfun$ type $\gamma$,\\
    $~\Gamma\,\vdash_{\textsf{4Dfun}}\,e\,:\,\CPonetwo{\tau}\langle\CPonetwo{\mu_1},\,(\CPonetwo{\alpha}\to\gamma)\rangle\,\gamma\,$\\
    $\langle\CPonetwo{\mu_2},\,(\CPonetwo{\beta}\to\gamma)\rangle\,\gamma$.
    \\
    (2) If 
    $~\Gamma\,\vdash_{\textsf{4Dfun}}\,e\,:\,\CPonetwo{\tau}\langle\CPonetwo{\mu_1},\,(\CPonetwo{\alpha}\to\gamma)\rangle\,\gamma\,$\\
    $\langle\CPonetwo{\mu_2},\,(\CPonetwo{\beta}\to\gamma)\rangle\,\gamma$
    for some $\sffDfun$ type $\gamma$, then\\
    $~\Gamma\,\vdash_{\textsf{CP}}\,e\,:\,\TrailMC{\tau}{\mu_1}{\alpha}{\mu_2}{\beta}$.
    \hfill$\qed$
\end{theorem}
The proof of this theorem is by induction on the typing derivation.
The first part of the theorem is formalized in Agda.
The second part requires that the term $e$ has a type that is the
image of $\CPonetwo{\cdot}$ for some $\gamma$.
We cannot simply write
$~\Gamma\,\vdash_{\textsf{4Dfun}}\,e\,:\,\TypTrailMc{\tau}{\mu_1}{\alpha\to\gamma}{\gamma}{\mu_2}{\beta\to\gamma}{\gamma}$,
because all the meta continuations that occurs in $\tau$, $\mu_1$,
$\alpha$, $\mu_2$, and $\beta$ must all use $\gamma$ consistently.
Currently, we do not know how to enforce this property in Agda.

As for the relationship between $\sffDfun$ and $\sffD$, 
the same discussion as what we have seen in part C.\ of \Sec{\ref{subsecSR}}
can be applied here.

\begin{theorem}[Typability between $\sffDfun$ and $\sffD$]
\label{thmfDtofunCP}
    ~\\
    (1) If $~\Gamma\,\vdash_{\textsf{4Dfun}}\,e\,:\,\TypTrailMc{\tau}{\mu_1}{\sigma_1}{\alpha}{\mu_2}{\sigma_2}{\beta}$, then \\
    $~\Gamma\,\vdash_{\textsf{4D}}\,e\,:\,\TypTrailMc{\tau'}{\mu'_1}{\sigma'_1}{\alpha'}{\mu'_2}{\sigma'_2}{\beta'}$ for some
    $\tau'$, $\mu'_1$, $\sigma'_1$, $\alpha'$, $\mu'_2$,
    $\sigma'_2$, and $\beta'$.\\
    (2) If $~\Gamma\,\vdash_{\textsf{4D}}\,e\,:\,\TypTrailMc{\tau}{\mu_1}{\sigma_1}{\alpha}{\mu_2}{\sigma_2}{\beta}$, then\\
    $~{\Gamma}\,\vdash_{\textsf{4Dfun}}\,e\,:\,{\tau'}
    \langle{\mu_1'},\,{\sigma_1'}\rangle\,$
    ${\alpha'} \langle{\mu_2'},\,{\sigma_2'}\rangle\,{\beta'}$ for some
    $\tau'$, $\mu'_1$, $\sigma'_1$, $\alpha'$, $\mu'_2$,
    $\sigma'_2$, and $\beta'$.
    \hfill$\qed$
\end{theorem}

%% file: other-SRZ.tex
\subsection{$\ttSRZ$}\label{subsecSRZ}

\begin{figure*}[h]
    \TypeSzRzMBOriginal\\[1ex]
    \TypeSzRzMBExtAbsSz
    \caption{$\sfMB$ type system \cite{Materzok2011Subtyping} and extended rules}
    \label{figTypeMB}
\end{figure*}


\begin{figure*}[h]
    \TypeSzRzMBTransFromMB\\[1em]
    \TypeSzRzMBTransToMB
    \caption{Type-level translation between $\sfMB$ and $\sffD'$}
    \label{figTypeMBTrans}
\end{figure*} 

\begin{figure*}[h]
    \TypeSzRzFromfDdash
    \caption{Type-level translation between $\sffD'$ and $\sffD$}
    \label{figTypefDdashTrans}
\end{figure*}

In this section, we compare the monomorphic version of 
Materzok and Biernacki's type system for $\ttSRZ$ 
(referred to below as $\sfMB$) \cite{Materzok2011Subtyping} with $\sffD$.
The relationship between $\sfMB$ and $\sffD$ is depicted as follows.
$\sfMB$ adopts a different notation called a type annotation to 
represent continuations and meta continuations, whose syntax
we introduce in the following paragraphs.
To connect $\sfMB$ and $\sffD$, we introduce $\sffD'$, which is the
same as $\sffD$ but without trails.
\begin{equation*}
\begin{tikzcd}[column sep=large]
\sfMB
    \arrow[r,"\textit{Meta continuation}",shift left=0.6ex]
  &[7ex]
\sffD'
    \arrow[r,"\textit{Add trail}", shift left=0.6ex]
    \arrow[l,"\textit{Type annotation}",shift left=0.6ex]
  &[2ex]
\sffD 
    \arrow[l,"\textit{Remove trail}", shift left=0.6ex]
\end{tikzcd}
\end{equation*}
We begin by introducing $\sfMB$'s calculus $\lambda_{S_0}$,
and then show that the typability of terms is preserved
between $\sfMB$ and $\sffD'$
and between $\sffD'$ and $\sffD$.

Below is the syntax of \lsz: a simply-typed $\lambda$-calculus with $\ttSRZ$.
\LambdaSZSyntaxOneColumn
The characteristic part of this language is the types and type annotations.
Type annotations show whether the term's surrounding context is empty $\epsilon$ or not.
The typing judgment goes
$\JudgeSTLC{\Gamma}{e}{\tau'_1\AnnMB{\tau_1}{\sigma_1}{\tau'_2}{\dots}\AnnMB{\tau_{n}}{\sigma_{n}}{\tau}{}}$,
and this reads
``under a typing environment $\Gamma$, an $\sfMB$ term $e$ is evaluated 
to a value of type $\tau$ in contexts of type
$\ArrowMB{\tau'_1}{\sigma_1}{\tau_1},\,\dots,\,\ArrowMB{\tau'_{n}}{\sigma_{n}}{\tau_{n}}$.''
This typing annotation can express contexts of any depth,
i.e., both continuations and meta continuations.
For example, 
$\JudgeSTLC{\Gamma}{e}{\tau_1\AnnMB{\tau_2}{\AnnMB{\tau_4}{\sigma_4}{\tau_5}{\sigma_5}}{\tau_3}{\sigma_3}}$ says that
``$e$ is evaluated with the continuation of type $\ArrowMB{\tau_1}{\AnnMB{\tau_4}{\sigma_4}{\tau_5}{\sigma_5}}{\tau_2}$ to the value of $\tau_3$.''
If the annotations $\sigma_4$ and $\sigma_5$ are both $\epsilon$,
the type corresponds to a meta continuation of type
$\tau_1\to(\tau_2\to\tau_4)\to\tau_5$.
Precise correspondence between $\sfMB$ types and $\sffD'$ types will be shown in
\Fig{\ref{figTypeMBTrans}}.

\Fig{\ref{figTypeMB}} shows the type system for $\lambda_{S_0}$.
The first five rules (from \Rule{MB-Var} to \Rule{MB-Reset0})
are taken from the previous work \cite{Materzok2011Subtyping}.
Instead of using the original type system, however, we replace
\Rule{MB-Abs} and \Rule{MB-Shift0} with \Rule{MB-Abs-Ext} and
\Rule{MB-Shift0-Ext}, respectively, also shown in \Fig{\ref{figTypeMB}}.
The rule \Rule{MB-Abs-Ext} is obtained by replacing $\sigma$
in \Rule{MB-Abs} with
$\AnnMB{\tau_3}{\sigma_3}{\tau_4}{\sigma_4}$.
Similarly, the rule \Rule{MB-Shift0-Ext} is obtained by instantiating
$\sigma_1$ and $\sigma_2$ in \Rule{MB-Shift0}
with non-empty annotations.

The typability of terms does not change with this replacement: the set
of terms that can be typed using \Rule{MB-Abs} and \Rule{MB-Shift0}
is the same as the one using \Rule{MB-Abs-Ext} and \Rule{MB-Shift0-Ext},
because none of the original rules in \Fig{\ref{figTypeMB}} requires
the annotation to be $\epsilon$\footnote{
We believe that the type annotation of the body of the functions and
$\ttSZ$ would never be empty, but we have not proved it yet.
}.

We are now ready to move on to the type-level translation between $\sfMB$ and $\sffD'$ in \Fig{\ref{figTypeMBTrans}}. 
The transformations $\leftharpoonup^{\tau}_{\textsf{D}'}$ and 
$\leftharpoonup^{\sigma}_{\textsf{D}'}$ convert types from $\sffD'$ to $\sfMB$.
Among these two translations, $\leftharpoonup^{\sigma}_{\textsf{D}'}$ 
takes a $\sffD'$ meta continuation and the $\sffD'$ answer type, 
then returns an $\sfMB$ type and an $\sfMB$ annotation.
The other pair of transformations $\rightharpoonup^{\tau}_{\textsf{D}'}$ and $\rightharpoonup^{\sigma}_{\textsf{D}'}$ 
is inversion of 
$\leftharpoonup^{\tau}_{\textsf{D}'}$ and $\leftharpoonup^{\sigma}_{\textsf{D}'}$ respectively, 
and satisfies the properties below.

\begin{lemma}\label{lemSRZFromToMB}
    For any $\lambda_{S_0}$ type $\tau_1$, $\toMBT{(\toFDT{\tau_1})} \equiv \tau_1$.
    Also, for any $\lambda_{S_0}$ annotation $\AnnMB{\tau_2}{\sigma_2}{\tau_3}{\sigma_3}$,\\
    $({(\toFDA{\tau_2}{\sigma_2})}^{\toMBAnn},\,{(\toFDA{\tau_3}{\sigma_3})}^{\toMBAnn})
    \equiv 
    ((\tau_2,\,\sigma_2),\,(\tau_3,\,\sigma_3))$.
    \hfill $\Box$
\end{lemma}

\begin{lemma}\label{lemSRZToFromMB}
    For any $\lambda_{D}$ type $\tau$,
    $\toFDT{(\toMBT{\tau})} \equiv \tau$.
    Also, for any $\lambda_{D}$ meta continuation type $\sigma_{\alpha}$ and 
    an answer type $\alpha$,\\
    ${(\toMBA{\sigma_{\alpha}}{\alpha})}^{\toFDAnn} \equiv (\sigma_{\alpha},\,\alpha)$. 
    \hfill $\Box$
\end{lemma}

Based on Lemmas \ref{lemSRZFromToMB} and \ref{lemSRZToFromMB},
an $\sfMB$ expression is also typable in $\sffD'$, and vice versa (Theorem \ref{thmFDMB}).
The proofs of these properties are by induction on typing derivations, 
and are formalized in Agda.

\begin{theorem}[Typability between $\sfMB$ and $\sffD'$]\label{thmFDMB}
    ~\\
(1) Suppose that
    $~\Gamma\,\vdash_{\sfMB}\,e\,:\,\tau_1\AnnMB{\tau_2}{\sigma_2}{\tau_3}{\sigma_3}$.
    If $\toFDT{\tau_1} = \tau$ and 
    $(\toFDA{\tau_2}{\sigma_2},\,\toFDA{\tau_3}{\sigma_3}) = ((\sigma_{\alpha},\,\alpha),\,(\sigma_{\beta},\,\beta))$
    holds, then\\
    $\toFDG{\Gamma}\,\vdash_{\sffD'}\,e\,:\,\TrailMC{\tau}{\sigma_{\alpha}}{\alpha}{\sigma_{\beta}}{\beta}$.
    \\
(2) Suppose that
    $~\Gamma'\,\vdash_{\sffD'}\,e'\,:\,\TrailMC{\tau'}{\sigma'_{\alpha}}{\alpha'}{\sigma'_{\beta}}{\beta'}$.
    If $\toMBT{\tau'} = \tau'_1$ and 
    $(\toMBA{\sigma'_{\alpha}}{\alpha'},\,\toMBA{\sigma'_{\beta}}{\beta'}) = ((\tau'_2,\,\sigma'_2),\,(\tau'_3,\,\sigma'_3))$ 
    holds, then
    $\toMBG{\Gamma'}\,\vdash_{\sfMB}\,e'\,:\,\tau'_1\AnnMB{\tau'_2}{\sigma'_2}{\tau'_3}{\sigma'_3}$.
    \hfill $\Box$
\end{theorem}

Lastly, we show the relationship between $\sffD'$ and $\sffD$.
$\sffD'$ is similar to $\sffD$ but without trails.
The type-level relationship between $\sffD'$ and $\sffD$ is shown in \Fig{\ref{figTypefDdashTrans}}.
Each translation from $\sffD'$ to $\sffD$, and from $\sffD$ to $\sffD'$ 
is defined as $\fDdashfDsuper{*}$ and  $\fDfDdashsuper{*}$, respectively.
The idea of these translations is similar to that of part B of
\Sec{\ref{subsecSR}}.
We simply add empty trails to $\sffD'$ type or remove trails from $\sffD$ type.
Based on these translations, we show that a $\sffD'$ term is typable in
$\sffD$, and vice versa.
The theorem below is proved by induction on the typing derivations, and it is formalized in Agda.
\newpage
\begin{theorem}[Typability between $\sffD'$ and $\sffD$]
    \label{thmSRZtrail}
    ~\\
    (1) If $~\Gamma\,\vdash_{\sffD'}\,e\,:\,\TrailMC{\tau}{\sigma_{\alpha}}{\alpha}{\sigma_{\beta}}{\beta}$, then \\
    $~\fDdashfDgamma{\Gamma}\,\vdash_{\sffD}\,e\,:\,\TypTrailMc{\fDdashfD{\tau}}{\bullet_{\mu}}{\fDdashfDmeta{\sigma_{\alpha}}}{\fDdashfD{\alpha}}{\bullet_{\mu}}{\fDdashfD{\sigma_{\beta}}}{\fDdashfD{\beta}}$.\\
    (2) If $~\Gamma\,\vdash_{\sffD}\,e\,:\,\TypTrailMc{\tau}{\mu_{\alpha}}{\sigma_{\alpha}}{\alpha}{\mu_{\beta}}{\sigma_{\beta}}{\beta}$, then \\
    $~\fDfDdashgamma{\Gamma}\,\vdash_{\sffD'}\,e\,:\,\TrailMC{\fDfDdash{\tau}}{\fDfDdashmeta{\sigma_{\alpha}}}{\fDfDdash{\alpha}}{\fDfDdashmeta{\sigma_{\beta}}}{\fDfDdash{\beta}}$.\hfill$\Box$
\end{theorem}

%% file: related.tex
\section{Related Work}\label{secRelated}

\paragraph{Typing Delimited Continuations}
There have been many type systems for delimited control operators.
As an extension of Danvy and Filinski's type system for $\ttSR$
\cite{Danvy1989Functional},
Asai and Kameyama \cite{Asai2007} show a type system for $\ttSR$
that supports let-polymorphism.
Materzok and Biernacki's type system for $\ttSRZ$
\cite{Materzok2011Subtyping} supports subtyping
and allows us to use captured continuations in different contexts.
Cong et al.~\cite{Cong2021Functional} and Kameyama and Yonezawa
\cite{Kameyama2008Typed} show different type systems for $\ttCP$.
(See Cong et al.~\cite{Cong2021Functional} for their comparison.)
In contrast to the previous work, the present paper shows a type
system for all the four delimited control operators, in particular,
$\ttCZ$.
On the other hand, our type system is monomorphic and does not support
any kind of polymorphism.

Dybvig, Peyton Jones, and Sabry~\cite{Dybvig2007Monadic} build a
monadic framework to type delimited continuations.
They present a new set of operators as building blocks to simulate the
existing operators such as 
$\ttS$ ($\Fpp{+}{+}$ in their paper), $\ttC$ ($\Fpp{+}{-}$), 
$\ttSZ$ ($\Fpp{-}{+}$), and $\ttCZ$ ($\Fpp{-}{-}$).
They show a Haskell type system, not for the four delimited control
operators per se, but for the set of basic operators they introduced.
Based on this work, Kiselyov \cite{kiselyov2010delimited} implements
four delimited control operators (in terms of the set of basic
operators) in OCaml which does not take answer types into account.

\paragraph{CPS Semantics for Delimited Control Operators}
The CPS interpreter for the four control operators in \Sec{\ref{secDsCps}}
is presented by Shan \cite{Shan2007Delimited}.
They show how $\ttSR$ simulates other control operators in an untyped
setting.
Biernacki, Danvy, and Millikin \cite{Biernacki2015Dynamic} 
present a CPS transformation that is equivalent to Shan's.
Their CPS transformation is derived from a defunctionalized version of
a definitional machine.

\paragraph{Algebraic Effects and Handlers}
Besides delimited control operators, 
algebraic effects and handlers \cite{plotkin2009handlers, plotkin2003algebraic} are another way to handle continuations.
The relationship between $\ttSZ$ and deep effect handlers 
has been established both under an untyped setting \cite{Forster2017Expressive} and a typed setting \cite{pirog2019typed, cong2022understanding}.


%% file: conclusion.tex
\section{Conclusion and Future Work}\label{secConclusion}

We have presented a monomorphic type system for four delimited control operators:
$\ttS$, $\ttC$, $\ttSZ$, and $\ttCZ$, from a corresponding CPS interpreter.
As far as we are aware of, we are the first to present the typing rule
for $\ttCZ$ that allows answer type modification
and to formalize the typing rules for all these operators in a unified notation.
Also, regarding the relationship between the previous studies,
we have shown that our type system subsumes the existing monomorphic
type systems.
 
One way to develop this work is to broaden the coverage of the type system
such as polymorphism and answer-type polymorphism.
Another approach is to understand the relationship 
between algebraic effect handlers using delimited control operators
under the typed setting.
As it is known that $\ttSZ$ and $\ttCZ$ are closely related to 
deep and shallow effect handlers respectively, 
this paper would become a solid foundation to explore their relationship.

%% file: appendix.tex
\section{Relationship between $\sffDfun$ and $\sffD$}

\begin{figure*}[h]
  \TypeLambdaVariablefDfun\\[0.5ex]
  \TypeSVariablefDfun\\[0.5ex]
  \TypePzVariablefDfun
  \caption{$\sffDfun$ type system for $\ttSR$}
  \label{figAppendixSRfDfun}
\end{figure*}

\begin{figure*}[h]
  \text{\fbox{$\JudgeSTLC{\Gamma}{e}{\TypTrailMc{\tau}{\bullet_\mu}{\sigma_{\alpha}}{\alpha}{\bullet_\mu}{\sigma_{\beta}}{\beta}}$}} \hfill
  \vspace{-1.6em}
  \TypeLambdaVariablefDdash\\[0.5ex]
  \TypeSVariablefDdash\\[0.5ex]
  \TypePzVariablefDdash
  \caption{$\sffD$ type system for $\ttSR$}
  \label{figAppendixSRfD}
\end{figure*}

In this section, we provide the full proof for 
the typability between $\sffDfun$ and $\sffD$, which we explained 
in \Thm{\ref{thmfDtofun1}} in part C of \Sec{\ref{secOtherTypeSystems}}.

The source languages for $\sffDfun$ and $\sffD$ both consist of 
a simply-typed $\lambda$ calculus extended with numbers and $\ttS$\\$\texttt{/reset}$.
The only difference is the type of the meta continuations:
in $\sffDfun$, the type is either an empty ($\bullet_{\sigma}$) or 
a function type ($\tau_1\to\tau_2$), 
whereas in $\sffD$, it is either an empty ($\bullet_{\sigma}$) or 
a product ($\Mc{\tau_1}{\mu_1}{\sigma_1}{\tau_2}{\mu_2}{\sigma_2}$). 
The typing rules of $\sffDfun$ and $\sffD$ are displayed in
Figures \ref{figAppendixSRfDfun} and \ref{figAppendixSRfD}, respectively.
Both type systems use \textsf{id-cont-type} in the rules of $\ttS$ and $\ttPZ$,
and its definition is the same as that of \Fig{\ref{figLDTypeSystem}}.

\subsection{Translation from $\sffDfun$ to $\sffD$}

This section corresponds to \Thm{\ref{thmfDtofun1}} (1).
Our approach is to use structural induction on the typing judgment of $\sffDfun$, 
which is described as
$~\Gamma\,\vdash_{\textsf{4Dfun}}\,e\,:\,\TypTrailMc{\tau}{\bullet_\mu}{\sigma_1}{\alpha}{\bullet_\mu}{\sigma_2}{\beta}$.

If $e$'s last used typing rule is one of 
\Rule{4Dfun-Var}, \Rule{4Dfun-Num}, \Rule{4Dfun-Lam} or \Rule{4Dfun-App}, 
it is obvious that if $e$ is typable in $\sffDfun$, 
then it is also typable in $\sffD$,
since each of the evaluation rules and the typing rules of $\sffDfun$
is identical to that of $\sffD$.

Next, if the last used typing rule is \Rule{4Dfun-Shift},
$e$ becomes $\Shift{k}{e}$, and its evaluation rule is as follows.
\[
  \begin{array}{l}
    \EfDfunCPS{\Shift{k}{e}}\,\rho\,
    \kappa\,=\,
    \Abs{t^{\Rmath{\mu_{\beta}^*}}}{}
    \Abs{m^{\Rmath{\sigma_{\beta}^*}}}{}\,\\
    \quad 
    \EfDfunCPS{e}\,\rho\,[
        \Abs{v^{\Rmath{\tau^*}}}{}
        \Abs{\kappa'^{\Rmath{(\Arrow{\tau_1}{\bullet_{\mu}}{\sigma_1}{\tau_2})^*}}}{}
        \Abs{t'^{\Rmath{\bullet_{\mu}^*}}}{}
        \Abs{m'^{\Rmath{\sigma_1^*}}}{}\\
    \quad\quad
        \kappa^{\Rmath{(\Arrow{\tau}{\mu_{\beta}}{\tau_1\to\tau_2}{\alpha})^*}}\,
        v^{\Rmath{\tau^*}}\,
        t^{\Rmath{\mu_{\beta}^*}}\,\\
    \quad\quad\quad
        (\Abs{v'^{\Rmath{\tau_1^*}}}{}
        \kappa'^{\Rmath{(\Arrow{\tau_1}{\bullet_{\mu}}{\sigma_1}{\tau_2})^*}}\,
        v'^{\Rmath{\tau_1^*}}\,
        t'^{\Rmath{\bullet_{\mu}^*}}\,
        m'^{\Rmath{\sigma_1^*}}
        )^{\Rmath{(\tau_1\to\tau_2)^*}}/k
    ]\\
    \hfill {k''_{id}}^{\Rmath{(\Arrow{\gamma}{\mu_{id}}{\sigma_{id}}{\gamma'})^*}}\,()^{\Rmath{\bullet_{\mu}^*}}\,
    m^{\Rmath{\sigma_{\beta}^*}}
  \end{array}
\]
At the first line, the interpreter receives $\kappa$ of type $(\Arrow{\tau}{\mu_{\beta}}{\tau_1\to\tau_2}{\alpha})^*$.
Also, at the third and the fourth line, 
$\kappa$ receives the meta continuation $(\Abs{v'}{\kappa'\,v'\,t'\,m'})$
of type $(\tau_1\to\tau_2)^*$.
Because the meta continuation has a form of function, the type of $t'$ and $m'$ do not explicitly appear on the rule \Rule{4Dfun-Shift}.
If we transform the meta continuation into defunctionalized form,
the evaluation rule becomes as follows.
\[
    \begin{array}{l}
    \ECPS{\Shift{k}{e}} \,\rho\,\kappa  \,=\,
    \Abs{t^{\Rmath{\mu_{\beta}^*}}}{}
    \Abs{m^{\Rmath{\sigma_{\beta}^*}}}{}\\
    \quad
    \ECPS{e}\,
    \rho[
        \Abs{v^{\Rmath{\tau^*}}}{}
        \Abs{\kappa'^{\Rmath{(\Arrow{\tau_1}{\bullet_{\mu}}{\sigma_1}{\tau_2})^*}}}{}
        \Abs{t'^{\Rmath{\bullet_{\mu}^*}}}{}
        \Abs{m'^{\Rmath{\sigma_1^*}}}{}\\
        \quad\quad\quad\quad
        \kappa^{\Rmath{(\Arrow{\tau}{\mu_{\beta}}{\Mc{\tau_1}{\bullet_{\mu}}{\sigma_1}{\tau_2}{\bullet_{\mu}}{\sigma_1}}{\alpha})^*}}
        v^{\Rmath{\tau^*}}\,
        t^{\Rmath{\mu_{\beta}^*}}\,\\
        \quad\quad\quad\quad\quad\quad\quad\quad
        ((\kappa'^{\Rmath{(\Arrow{\tau_1}{\bullet_{\mu}}{\sigma_1}{\tau_2})^*}},\,t'^{\Rmath{\bullet_{\mu}^*}}) :: m'^{\Rmath{\sigma_1^*}})
        / k
    ] \,\\
    \hfill
    \idk^{\Rmath{(\Arrow{\gamma}{\mu_{id}}{\sigma_{id}}{\gamma'})^*}}\,\,
    ()^{\Rmath{\bullet_{\mu}^*}} \,\,
    m^{\Rmath{\sigma_{\beta}^*}}
    \end{array}
\]
The meta continuation consists of the same parameters $\kappa'$, $t'$, and $m'$,
and each of their types is also the same as the previous evaluation rule.
Therefore, if $\Shift{e}{k}$ is typed using \Rule{4Dfun-Shift}, 
it is also typed using \Rule{TShift$'$}.

Finally, if the last used typing rule is \Rule{4Dfun-Prompt0}, 
$e$ becomes $\myReset{e}$.
Below are the two interpreters from $\sffDfun$ and $\sffD$.
\[
    \begin{array}{l}
      \EfDfunCPS{\myReset{e}} \,\rho\,\kappa^{\Rmath{(\Arrow{\tau}{\bullet_{\mu}}{\sigma_{\alpha}}{\alpha})^*}}  \, =\,
      \Abs{t^{\Rmath{\bullet_{\mu}^*}}}{}
      \Abs{m^{\Rmath{\sigma_{\alpha}^*}}}{}\,\\
      \quad\quad\quad
      \EfDfunCPS{e}\,\rho\,
      \idk^{\Rmath{(\Arrow{\gamma}{\mu_{id}}{\sigma_{id}}{\gamma'})^*}}\,
      ()^{\Rmath{\bullet_{\mu}^*}}\,\\
      \quad\quad\quad\quad\quad\quad\quad\quad\quad
      (\Abs{v^{\Rmath{\tau^*}}}{}
      \kappa^{\Rmath{(\Arrow{\tau}{\bullet_{\mu}}{\sigma_{\alpha}}{\alpha})^*}}\,
      v^{\Rmath{\tau^*}}\,
      t^{\Rmath{\bullet_{\mu}^*}}\,
      m^{\Rmath{\sigma_{\alpha}^*}})
      \\[1ex]
      \ECPS{\myReset{e}} \,\rho\,\kappa^{\Rmath{(\Arrow{\tau}{\bullet_{\mu}}{\sigma_{\alpha}}{\alpha})^*}}  \, =\,
      \Abs{t^{\Rmath{\bullet_{\mu}^*}}}{}
      \Abs{m^{\Rmath{\sigma_{\alpha}^*}}}{}\,\\
      \quad\quad\quad
      \ECPS{e}\,\rho\,
      \idk^{\Rmath{(\Arrow{\gamma}{\mu_{id}}{\sigma_{id}}{\gamma'})^*}}\,
      ()^{\Rmath{\bullet_{\mu}^*}}\,\\
      \hfill
      ((\kappa^{\Rmath{(\Arrow{\tau}{\bullet_{\mu}}{\sigma_{\alpha}}{\alpha})^*}},\,
      t^{\Rmath{\bullet_{\mu}^*}}),\,
      m^{\Rmath{\sigma_{\alpha}^*}})
    \end{array}
\]
The difference between these two is the form of meta continuations 
$(\Abs{v}{\kappa\,v\,t\,m})$ at the third line, 
and $((\kappa,\,t),\,m)$ at the sixth line.
Both meta continuations have the same parameters $\kappa$, $t$, and $m$,
and each of their types is the same.
Therefore, if $\myReset{e}$ is typed using \Rule{4Dfun-Prompt0}, 
it is also typed using \Rule{TPrompt0$'$}.

\subsection{Translation from $\sffD$ to $\sffDfun$}

This section corresponds to the \Thm{\ref{thmfDtofun1}} (2), 
and the proof strategy is the same as in the previous section.
We use structural induction on the typing judgment of $\sffD$, 
which is described as
$~\Gamma\,\vdash_{\textsf{4D}}\,e\,:\,\TypTrailMc{\tau}{\bullet_\mu}{\sigma_1}{\alpha}{\bullet_\mu}{\sigma_2}{\beta}$.

From \Rule{TVar$'$} to \Rule{TApp$'$}, it is obvious because 
their evaluation rules and the typing rules are identical.

In the case of \Rule{TShift$'$} and \Rule{TPrompt0$'$}, 
the same argument of the previous section applies here.
Because both interpreters have the same parameters, and each of them has the same type,
if $e$ is typed in $\sffD$, it is typed in $\sffDfun$.

